\documentclass[fleqn,usenatbib]{mnras}

\usepackage[T1]{fontenc}

\DeclareRobustCommand{\VAN}[3]{#2}
\let\VANthebibliography\thebibliography
\def\thebibliography{\DeclareRobustCommand{\VAN}[3]{##3}\VANthebibliography}

\usepackage{graphicx}	
\usepackage{amsmath}	
\usepackage{amssymb}	

\usepackage{newtxtext,newtxmath}

\title[SuperWASP Variable Stars]{SuperWASP Variable Stars: Classifying Light Curves Using Citizen Science}

\author[H. B. Thiemann et al.]{
Heidi B. Thiemann,$^1{}^2$\thanks{E-mail: heidi.thiemann@open.ac.uk (HBT)}
Andrew J. Norton,$^{1}$
Hugh J. Dickinson,$^{1}$
Adam McMaster$^1{}^2$
\newauthor
Ulrich C. Kolb,$^{1}$
\\
$^{1}$School of Physical Sciences, The Open University, Milton Keynes, MK7 6AA, UK\\
$^{2}$DISCnet Centre for Doctoral Training, The Open University, Walton Hall, Milton Keynes, MK7 6AA, UK
}

\date{Accepted 2021 January 14. Received 2021 January 14; in original form 2020 December 10}

\pubyear{2021}

\begin{document}
\label{firstpage}
\pagerange{\pageref{firstpage}--\pageref{lastpage}}
\maketitle

\begin{abstract}

We present the first analysis of results from the SuperWASP Variable Stars Zooniverse project, which is aiming to classify 1.6 million phase-folded light curves of candidate stellar variables observed by the SuperWASP all sky survey with periods detected in the \textit{SuperWASP periodicity catalogue}. The resultant data set currently contains $>$1 million classifications corresponding to $>$500,000 object-period combinations, provided by citizen scientist volunteers. Volunteer-classified light curves have $\sim$89 per cent accuracy for detached and semi-detached eclipsing binaries, but only $\sim$9 per cent accuracy for rotationally modulated variables, based on known objects. We demonstrate that this Zooniverse project will be valuable for both population studies of individual variable types and the identification of stellar variables for follow up. We present preliminary findings on various unique and extreme variables in this analysis, including long period contact binaries and binaries near the short-period cutoff, and we identify 301 previously unknown binaries and pulsators. We are now in the process of developing a web portal to enable other researchers to access the outputs of the SuperWASP Variable Stars project.

\end{abstract}




\section{Introduction}

Variable stars are key to investigating and testing stellar astrophysics, and the dynamics and structure of stellar systems. The detection, classification, and study of classes of variable stars is therefore an important pursuit. Typically, variable stars are detected through amplitude and period variations in their photometric light curve. Classifications of periodic variables based on their light curve are not always conclusive, but instead give a strong indication of variable type, and can be used to identify candidates for spectroscopic and photometric follow-up.

The full SuperWASP photometric archive contains $>$30 million light curves of relatively bright stars (V$\leq$15), observed with a high cadence (as short as 30 seconds) and long baseline ($\sim$11 years). A previous period search using the first few years of the SuperWASP archive enabled a significant amount of research in the field of stellar variability, including: the identification of 140 short-period eclipsing binaries close to the period cut-off \citep{lohr2013}; the identification of period change in post common-envelope eclipsing binary systems to search for circumbinary planets \citep{lohr2014}; the discovery of a doubly eclipsing quintuple system \citep{lohr2015}; the identification of period change in $\sim$1400 eclipsing binaries \citep{lohr2015b}; the discovery of a $\delta$ Sct star in an eclipsing binary \citep{norton2016}; the study of $\sim$5000 RR Lyrae stars and identification of $\sim$800 Blazhko effect systems \citep{greer2017}; and the study of rotationally modulated variables \citep{thiemann2020}. A more recent re-analysis of this archive detected $\sim$8 million potential periods in $\sim$3 million unique objects \citep{norton2018}.

There have been previous attempts at using machine learning algorithms and Artificial Neural Networks (ANNs), often called Neural Networks (NN), to automate the classification of SuperWASP variable stars from the raw data, including \citet{payne2013}, who made use of three NNs to process a range of parameters which defined the shape of the phase folded light curve. They processed over 4.3 million periods, giving $\sim$1.1 million preliminary classifications. However these NNs found only partial success, identifying 75 per cent of light curves correctly. As an alternative to machine learning, the SuperWASP Variable Stars (SVS) Zooniverse\footnote{\url{www.zooniverse.org/projects/ajnorton/superwasp-variable-stars}} project is instead using citizen science to classify the 1.6 million folded light curves referred to above. In this paper, we present the first analysis of SVS, containing over 1 million classifications, corresponding to over 500,000 unique object-period combinations. 

\begin{figure}
 \includegraphics[width=\columnwidth]{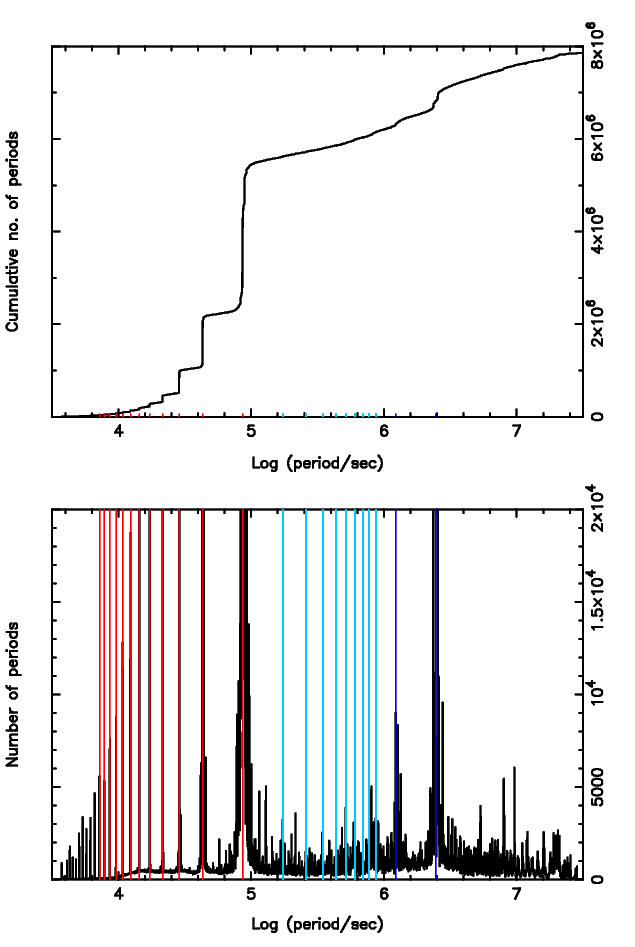}
 \caption{Histogram of the identified periods in all objects in the \textit{SuperWASP Periodicity Catalogue}. There are significant numbers of excess periods close to integer multiples or fractions of a sidereal day or lunar month, indicated by coloured vertical lines (red lines correspond to fractions of a day; light blue corresponds to multiples of a day; dark blue corresponds to the monthly and linger cycles). All such periods are flagged and may be discarded. The upper panel shows the cumulative period histogram while the lower one, whose vertical axis is truncated, shows the regular histogram.}
 \label{fig:hist_allflags}
\end{figure}

\begin{figure}
 \includegraphics[width=\columnwidth]{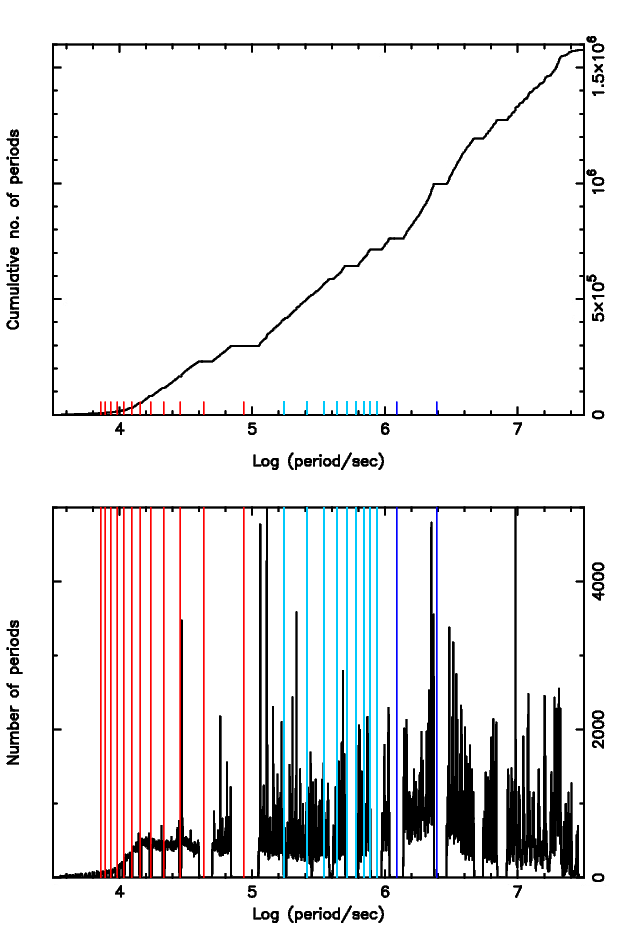}
 \caption{Histogram of all un-flagged periods corresponding to objects in the \textit{SuperWASP Periodicity Catalogue}. The coloured vertical lines indicate where flagged periods have been removed (red lines correspond to fractions of a day; light blue corresponds to multiples of a day; dark blue corresponds to the monthly and linger cycles). The upper panel shows the cumulative period histogram while the lower one shows the regular histogram.}
 \label{fig:hist_zeroflags}
\end{figure}

The SVS project was launched on 5th Sep 2018 and had engaged $\sim$4,500 volunteers at the time of this analysis. This analysis acts as a preliminary look at the Zooniverse classifications, demonstrating that SVS can be used for both population studies and for identifying rare and unique variables. This analysis will guide how we develop the project as it gains more volunteer and machine learning classifications. In Section \ref{sec:data} we describe the SuperWASP data; in Section \ref{sec:citsci} we describe the Zooniverse project; in Section \ref{sec:results} we summarise our results including the identification of new and unique stellar variables; in Section \ref{sec:conclusions} we draw our conclusions.

\section{SuperWASP Periodicity Catalogue}
\label{sec:data}

SuperWASP \citep{Pollacco2006} surveyed almost the entire night sky using two identical observatories in La Palma, Canary Islands, and Sutherland, South Africa. Each robotic observatory consisted of 8 cameras each with a 14 cm aperture and a 7.8 $\times$ 7.8 square degree field of view, allowing for a total sky coverage of $\sim$500 square degrees per exposure. The survey excludes the Galactic Plane where the large pixel scale of 16.7 arcsecond per pixel prevents separation of signals from individual stars in this dense stellar region. SuperWASP observations were reduced using the pipeline described in \citet{Pollacco2006}. Over the course of $\sim$2800 nights between 2004 - 2013, SuperWASP accumulated $\sim$16 million images containing $\sim$580 billion data points corresponding to $\sim$31 million unique stars \citep{norton2018}. The SuperWASP data set therefore provides a high cadence and long baseline of observations for more than 30 million stars with magnitudes between $V=8-15$.

For SuperWASP observations, 1 count $s^{-1}$ after background subtraction is roughly equivalent to V$\sim$15. Therefore the mean SuperWASP magnitude is defined as $V = -2.5 \log_{10}(\frac{F}{10^6})$ where $F$ is the mean SuperWASP flux and the pseudo-V magnitude is comparable to the Tycho V magnitude. A typical object in the SuperWASP archive will have $\sim$20,000 observations in its light curve. While the SuperWASP data can contain a significant level of noise, the long baseline of observations can often compensate for this in phase folded light curves. 

SuperWASP photometry is carried out by placing apertures on the images at pre-defined positions identified using the USNO catalogue as an input. However, the large pixel size of the individual cameras means that it is possible that a single star can be associated with two or more different identifiers in the SuperWASP archive, and that light from multiple stars can appear within the same photometric aperture. Typically there is only a single (or dominant) star in the aperture, so association with a specific object is possible, but that is not always the case. Hence, in each case confirmatory photometry with a small PSF is necessary to confirm exactly which object is variable.

\begin{figure}
 \includegraphics[width=\columnwidth]{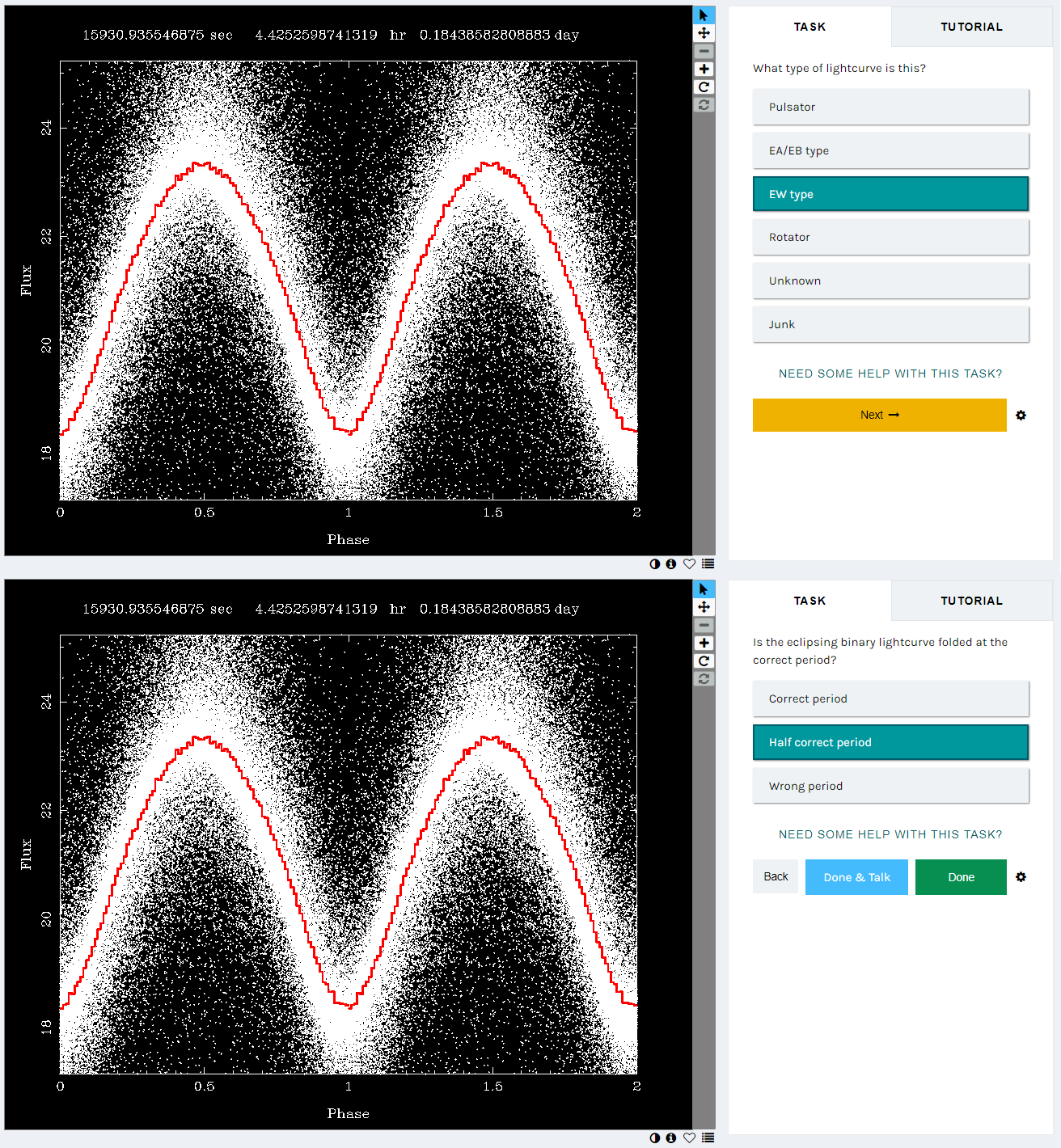}
 \caption{\textbf{Upper panel:} Volunteers are first tasked with classifying each light curve as a generic variable type. This example shows an EW folded at half the correct period. \textbf{Lower panel:} If a volunteer chooses a classification of EA/EB, EW, or pulsator, they are asked to choose whether the period is correct or not.}
 \label{fig:EW_first}
\end{figure}

\citet{norton2018} recently performed a re-analysis of the entire SuperWASP archive with the aim of detecting all periodic variables. The re-analysis comprised a one-dimensional CLEAN power spectrum analysis (based on the technique outlined by \citet{roberts1987}) as well as a phase dispersion minimisation and folding analysis (following the method of \citet{davies1990}). Only periods that were significantly detected using \textit{both} methods were considered to be plausible. For each light curve, all periods that passed these criteria were recorded, with a significance value recorded from both the folding analysis and the Fourier analysis. The periods identified have an average uncertainty of $\sim\pm0.1$ per cent.

This re-analysis detected $\sim$8 million candidate periods of stellar variables in $\sim$3 million unique objects, shown in Figure \ref{fig:hist_allflags}. A significant number of period detections result from systematic effects in the SuperWASP photometric data, resulting in the detection of periods close to integer fractions or multiples of a sidereal day or lunar month (i.e. 1 day, 1/2 day, 1/4 day, etc.). Periods flagged as affected by one of these effects were removed from the data set, leaving 1,569,061 candidate periods in 767,199 unique objects, shown in Figure \ref{fig:hist_zeroflags}. Clearly some genuine periods will have been rejected by this method, but if we extrapolate across the gaps, the rejected genuine periods should amount to no more than 5 per cent of the total. The \textit{SuperWASP periodicity catalogue} is available on the Warwick SuperWASP archive\footnote{\url{http://wasp.warwick.ac.uk/archive/docs/index.shtml}} as the table \textit{period\_ajn5}.

To generate subjects for SVS \citep{norton2018}, light curve data for objects with one or more potentially genuine periods listed in the \textit{SuperWASP Periodicity Catalogue} were used. The data for each selected object were folded at each of its potential periods and then rendered to produce a set of one or more phase-folded light curve images. Each image displays the raw photometric data points, overlaid with the mean profile in 100 phase bins, an example of which is shown in Figure \ref{fig:EW_first}.

\section{Citizen Science}
\label{sec:citsci}

The Zooniverse\footnote{\url{www.zooniverse.org}} \citep{lintott2008} is the world's most popular platform for "people-powered research", where a community of volunteers, or "citizen scientists", can participate in real scientific research through simple tasks such as analysing and categorising large data sets. This approach, using the "wisdom of the crowd", can be used to greatly improve the accuracy and speed with which data can be analysed and classified. Despite minimal training and subject matter expertise, Zooniverse volunteers have proven time and time again that non-experts can achieve a good level of accuracy, and can identify unusual objects that automated algorithms will often miss.

SVS launched on 5th September 2018, with the aim of classifying the output of the \textit{SuperWASP Periodicity Catalogue} \citep{norton2018}. The aim of SVS is threefold: to identify rare variable stars; to identify populations of variable stars in order to probe the extremes and trends of each population; and to facilitate the future development of a web portal in order to give researchers and the public access to the output of this project. 

We constructed the SVS project using the Zooniverse project builder platform\footnote{\url{www.zooniverse.org/lab}}, creating a classification task, tutorial, and "Field Guide" which provides example light curves and guidance for classification. There is also an option for volunteers to report their findings in the "Talk" section, where they can discuss individual light curves, highlight unusual and rare ones, and identify which objects have already been detected in other databases.

The classification of variable stars can be difficult, with 211 variable star types and sub-types listed in the International Variable Star Index\footnote{\url{www.aavso.org/vsx/index.php}} (VSX) \citep{2015watson}. The noise level of the SuperWASP light curves often makes it difficult to distinguish between similar types of variables. However, to be successful, Zooniverse, projects must be accessible to non-subject matter experts. We therefore ask volunteers to classify light curves into the following generic and overarching variable types: 

\begin{itemize}
    \item Pulsators: stars which display periodic changes in brightness due to changes in the star's size and luminosity as its outer layers expand and contract in a regular manner. This category includes RR Lyrae, $\delta$ Scuti, Cepheid variables, and Mira variables. Light curves are often asymmetric with a steeper rise and shallower fall in brightness.
    \item EA/EB: detached and semi-detached eclipsing binary systems which display periodic changes in brightness. This category includes Algol (EA) and Beta Lyrae (EB) eclipsing binaries. Two eclipses per cycle may be distinguished, often of different depth, with clear boundaries to the eclipses.
    \item EW: contact-eclipsing and near-contact eclipsing binary systems which display periodic changes in brightness. This category includes W Ursae Majoris (EW) type eclipsing binaries. Brightness variation is continuous and the eclipses are often of similar depth, resulting in half the orbital period often being identified instead of the true period.
    \item Rotators: stars which display rotational modulation in their light curve. This category includes single stars with significant star spots and stars with ellipsoidal modulation from close binaries that do not eclipse but instead are distorted into non-spherical (ellipsoidal) shapes by gravity due to their proximity. Brightness variations are typically quasi-sinusoidal.
    \item Unknown: stars displaying some degree of periodicity but which do not fall into any previous category. This category might include semi-regular stars and long period variables.
    \item Junk: light curves which display no genuine periodicity, or apparent periodicity which is due only to data dropouts or remaining systematic artefacts.
\end{itemize}

Volunteers are presented with a phase-folded light curve and tasked with classifying it into one of the following options: pulsator, EA/EB, EW, rotator, unknown, or junk, shown in Figure \ref{fig:EW_first}. If the volunteer chooses either EA/EB, EW, or pulsator, they are presented with a second question which asks them to choose whether the folding period is: correct period, half period, or wrong period. The classification task itself is essentially a pattern matching task.

We collect multiple classifications of each phase-folded light curve, allowing us to take the most common classification as the true classification and "retire" it from the live project. Between 5th September 2018 -- 23rd September 2019, each light curve required 7 classifications from separate volunteers to "retire" it, meaning that if a light curve received 4 or more of the same classification, the light curve would be assigned to the corresponding category. On 24th September 2019, a variable retirement rate was implemented using Caesar\footnote{\url{https://caesar.zooniverse.org}} advanced retirement engine provided by the Zooniverse platform. As a result, a light curve is now retired if either the classification count reaches 7, the subject receives 4 of the same classification, or if the subject receives 3 junk classifications, since junk light curves are typically easier to identify. Following the introduction of the variable retirement rate with Caesar, junk classified subjects are retired more quickly, so we would expect to see a higher relative frequency of junk in the output, with the number of junk classifications eventually plateauing as they are retired from the live project.

In the period immediately following the project launch, the subject images presented to volunteers were selected randomly from the full pool of 1.6 million light curves. Even if all 4,500 volunteers that had so far engaged with the project classified one subject per minute, the expected time for any particular subject to accrue 7 classifications is almost 40 hours. In reality, the initial retirement rate was $\sim$3,000 subjects per month on average. Accordingly, a subject batching strategy was adopted which reduced the available subject pool size to 288,000 light curves at any one time. Following this change, the retirement rate increased to $\sim$17,000 subjects per month, peaking at $\sim$43,711 retirements in October 2019. 

During peak times of activity (when SVS is promoted as a "featured project" on the Zooniverse front page), there is an average of $\sim$4,300 classifications per day, peaking at 11,442; outside of these intervals, there is an average of $\sim$1,100 classifications per day and a retirement rate of $\sim$5,000 per month. At this lower classification rate, it is estimated that it will take $\sim$4--5 years to complete each "live" set of 288,000 objects, or $\sim$25 years to complete the full set ($\sim$15 years at a higher classification rate). By comparison, one of the authors classified $\sim$5,000 light curves in a day without working on other research activities. Considering these timescales, machine learning will be vital to complete the classification of all 1.6 million phase-folded light curves within a reasonable time-frame.

We use the Gini coefficient to give a quantitative measure of the engagement of volunteers. The Gini coefficient ranges from 0 to 1, where 0 indicates that each volunteer contributes an equal number of classifications, and 1 indicates that there is an extreme difference in number of classifications from each volunteer. We find that the mean Gini coefficient for SVS is 0.92. By comparison, \citet{gini} finds that the mean Gini coefficient for astronomy projects on Zooniverse is 0.82, and \citet{Eisner} finds a similarly high Gini coefficient for \textit{Planet Hunters TESS} of 0.94. Whilst a higher Gini coefficient does not necessarily indicate project "success", it does indicate that SVS has a large number of prolific classifiers, which is often desirable for citizen science projects. Loyal classifiers spend more time engaging with the project, and hence are likely to have a strong understanding of the project aims and classification methods and make fewer mistakes.

For the project age, SVS has fewer total volunteers than other general astronomy projects on the Zooniverse, but a comparable number of total volunteers to other non-astronomy projects and variable star astronomy projects. A direct comparison is \textit{Variable Star Zoo}\footnote{\url{https://www.zooniverse.org/projects/ilacerna/variable-star-zoo}} (classifying $\sim$60,000 light curves), a project which aims to classify variable stars in the VVV survey. \textit{Variable Star Zoo} launched in July 2018 and has engaged with 5,305 volunteers to date, similar to SVS. Two upcoming variable star Zooniverse projects are \textit{Zwicky Stellar Sleuths}\footnote{\url{https://www.zooniverse.org/projects/adamamiller/zwickys-stellar-sleuths}}, and a new project by ASAS-SN, \textit{Citizen ASAS-SN}\footnote{\url{https://www.zooniverse.org/projects/tharinduj/citizen-asas-sn}}. SVS will complement these projects, and the increase in variable star Zooniverse projects may increase volunteer interest in this branch of astronomy.

\begin{figure}
 \includegraphics[width=\columnwidth]{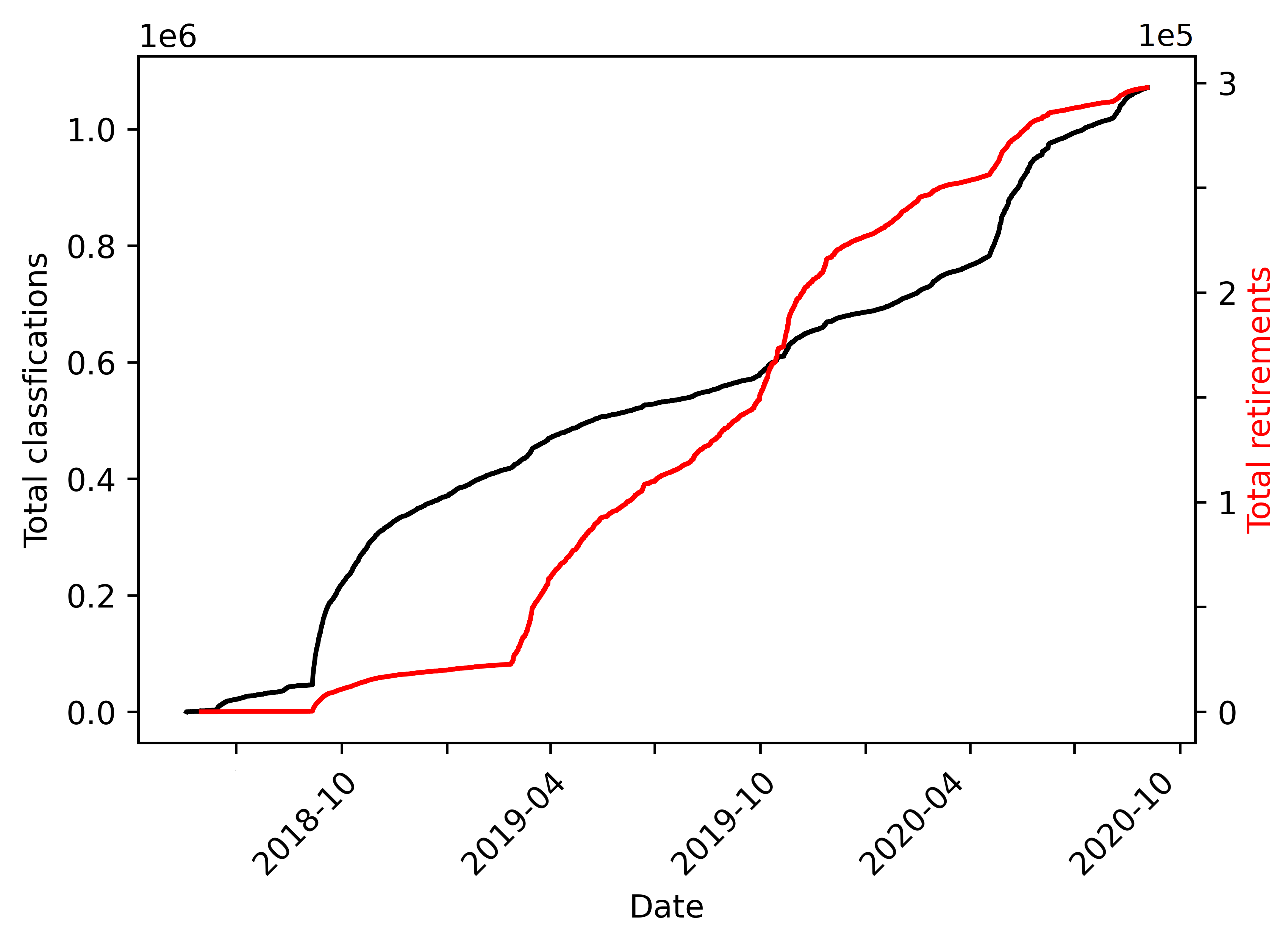}
 \caption{The number of classifications (black) and retirements (red) over the first 2 years of the project. The shallow increase shows pre-launch classifications from experts and beta testers. SVS was officially launched on 2nd September 2018, and since then has has a fairly consistent classification rate. Peaks of activity (such as being a "featured project") cause sudden rises in classifications. The change to a variable retirement limit and batching is clear in early 2019.}
 \label{fig:class_timeseries}
\end{figure}

\subsection{Data Cleaning}
\label{sec:data_cleaning}

The classifications used in this analysis were downloaded on 2nd September 2020, giving almost 2 years of classification data. Although there have been 1,071,345 classifications corresponding to over 568,739 unique object-period combinations, the majority of light curves have not yet received a sufficient number of classifications for retirement. 

Classifications from SVS are exported as a CSV file from the Zooniverse site. Before data cleaning, the SVS classification export is stripped of non-essential data, including time of classification and username of Zooniverse volunteers. In addition to the primary science analysis, an in-depth assessment of classification reliability, including detection of "spam" classifications was performed. For this secondary analysis, the full SVS classification export was used as is.

\begin{figure}
 \includegraphics[width=\columnwidth]{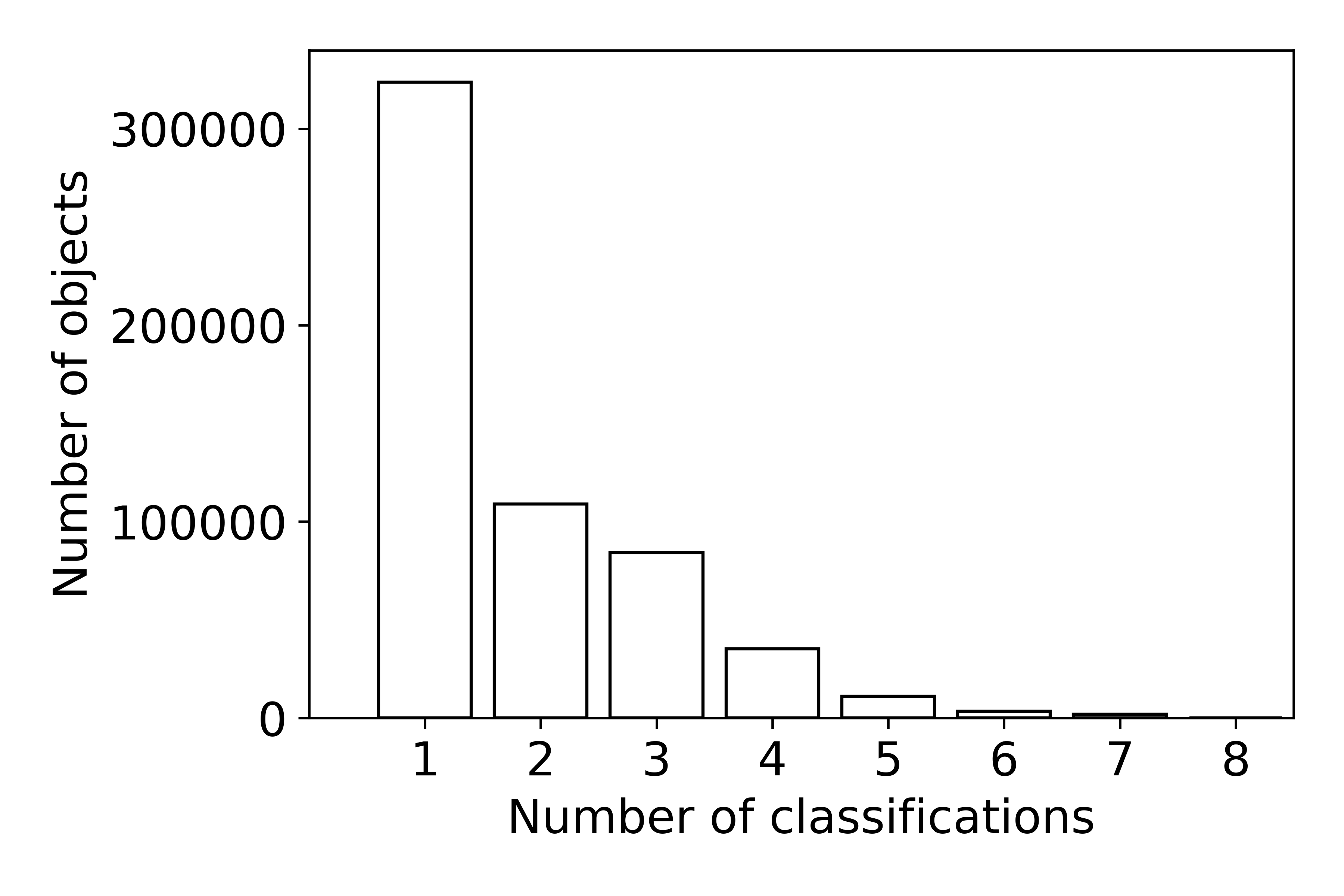}
 \caption{There are 5 objects with 9 classifications, 27 objects with 8 classifications, 1934 objects with 7 classifications, and 3510 objects with 6 classifications, 11,085 with 5 classifications, 35,298 with 4 classifications, 84,180 with 3 classifications, 109,034 with 2 classifications, 323,666 with 1 classification. At this stage, only 9 per cent of objects (7 per cent of non-junk objects) have received enough classifications for retirement.}
 \label{fig:number}
\end{figure}

The likely classification for each subject is decided by a custom written script. This script looks at all the classifications of the same Subject ID (or same SuperWASP ID and Period ID) and finds the most popular (or only) classification. If two (or more) classifications are equally popular, then we allocate the classification as the first given classification from the following list: junk, pulsator, rotator, EW, EA/EB, unknown (ordered from most common to least common). The unfiltered SVS export has 1,071,345 rows corresponding to all classifications made up to that time. After processing and removing duplicated rows, 1,025,750 light curve classifications remain. After finding the top classification for each subject, the output had 568,739 rows corresponding to unique object-period combinations. Figure \ref{fig:number} shows a histogram of the number of classifications per object.

Additional catalogues are cross-matched with the output to identify additional parameters such as distance, colour, and previous classifications. This includes a 10 arcsecond spatial cross-match with Gaia-DR2 and the Gaia-DR2 Bailer-Jones distance catalogue (\citealt{GaiaB,bailer2018}), a 10 arcsecond cross-match with NOMAD \citep{nomad}, and a 2 arcminute cross-match to VSX \citep{2015watson}. 

Light curves with fewer than 4 classifications are removed, and any remaining duplicates (both spatial and WASP ID) are retained, since these are plausibly multi-periodic or multi-classification objects. We complete an initial visual assessment of unrealistic periods, but at this stage, objects with such periods are not removed since these are plausibly extreme period objects which may be of interest. Table \ref{tab:initial_data_cleaning} shows a breakdown of the cleaned data set.

\begin{table*}
 \caption{Breakdown of the first 1 million classifications corresponding to 568,739 unique object-period combinations, and the results of positional cross-matches to the Gaia-DR2 and \citet{bailer2018} catalogue, VSX, and SuperWASP catalogues of binaries \citep{payne2013} and pulsators \citep{greer2017}.} 
 \label{tab:initial_data_cleaning}
 \begin{tabular}{lccccccc}
  \hline
   & Full output & EA/EB & EW & Pulsator & Rotator & Unknown & Junk \\
  \hline
   Classifications & 568739 & 29882 & 36328 & 25730 & 56582 & 41541 & 378,671 \\
   $N_{class} \geq 4$ & 13390 & 2425 & 3187 & 1777 & 4402 & 1599 & N/A \\
   $N_{class} \geq 4$ and correct period & 11322 & 1629 & 2672 & 1020 & 4402 & 1599 & N/A \\
   In Gaia-DR2 & 10213 & 792 & 2599 & 1000 & 4275 & 1547 & N/A \\
   In VSX & 5,283 & 665 & 1528 & 579 & 1939 & 572 & N/A \\
   In Payne and/or Greer & 314 & 259 & 44 & 11 & N/A & N/A & N/A \\
  \hline
 \end{tabular}
\end{table*}

\subsection{Classification Reliability}
\label{sec:assessing}

\begin{figure}
 \includegraphics[width=\columnwidth]{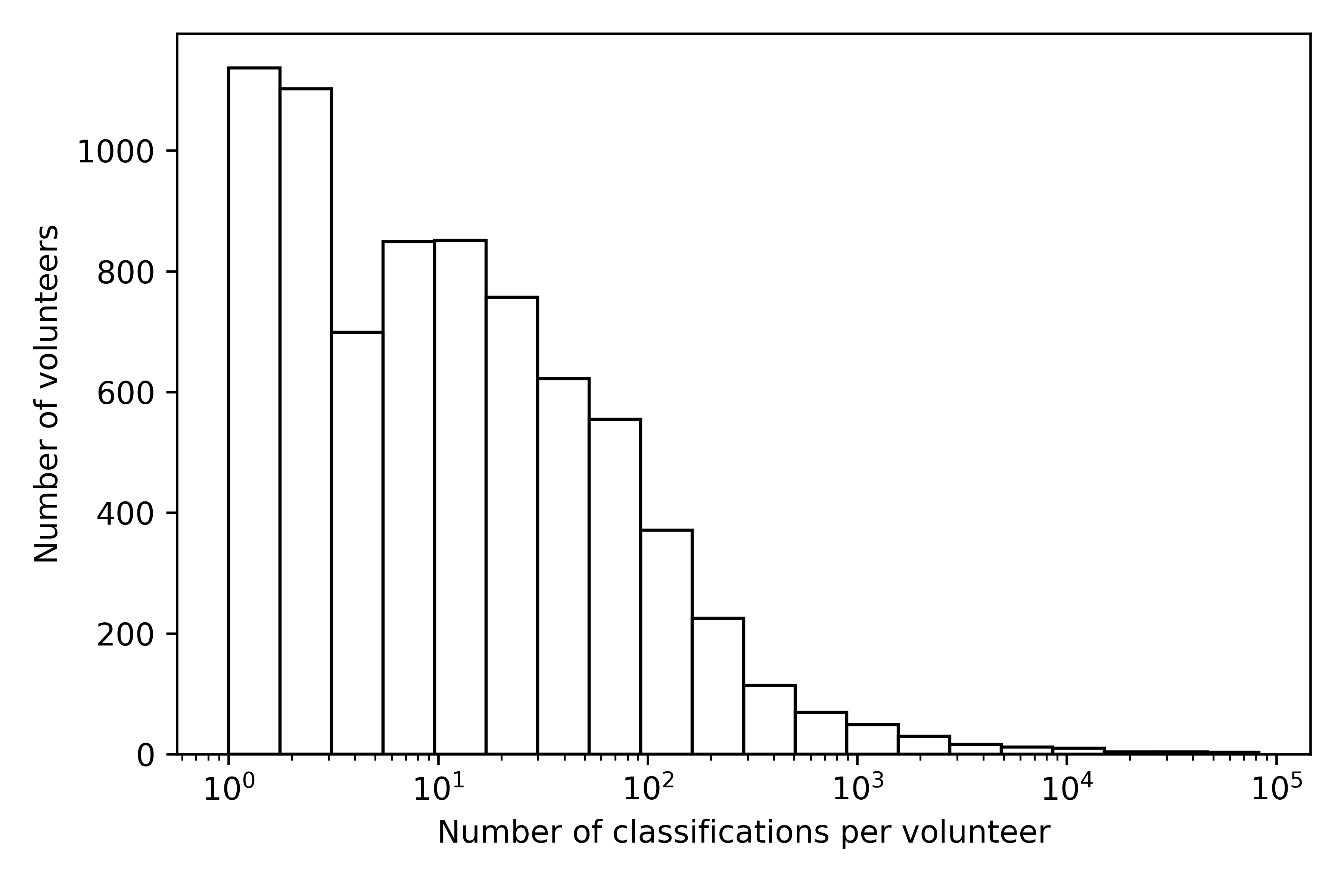}
 \caption{The number of classifications per volunteer. Any classifications made by an anonymous volunteer over different days will be counted as multiple volunteers' inputs.}
 \label{fig:volunteer_nclassification}
\end{figure}

A total of 7,478 volunteers made 1,071,345 classifications. SVS has $\sim$4,500 registered volunteers, indicating that $\sim$3000 volunteers engaged with the project but did not register on the Zooniverse platform. Registered volunteers made 93.9 per cent of classification, and 6.1 per cent of classifications (65,398) were made by unregistered or anonymous volunteers, making $\sim$20 classification each on average. Fig \ref{fig:volunteer_nclassification} shows the distribution of classifications made per volunteer. Just over half (52.6 per cent) of volunteers made 10 or fewer classifications, 36.0 per cent made 11--100, 9.6 per cent made 101--1000, and 1.6 per cent made over 1,000. 18 (0.2 per cent) "super-classifiers" made more than 10,000 classifications.

\begin{figure}
 \includegraphics[width=\columnwidth]{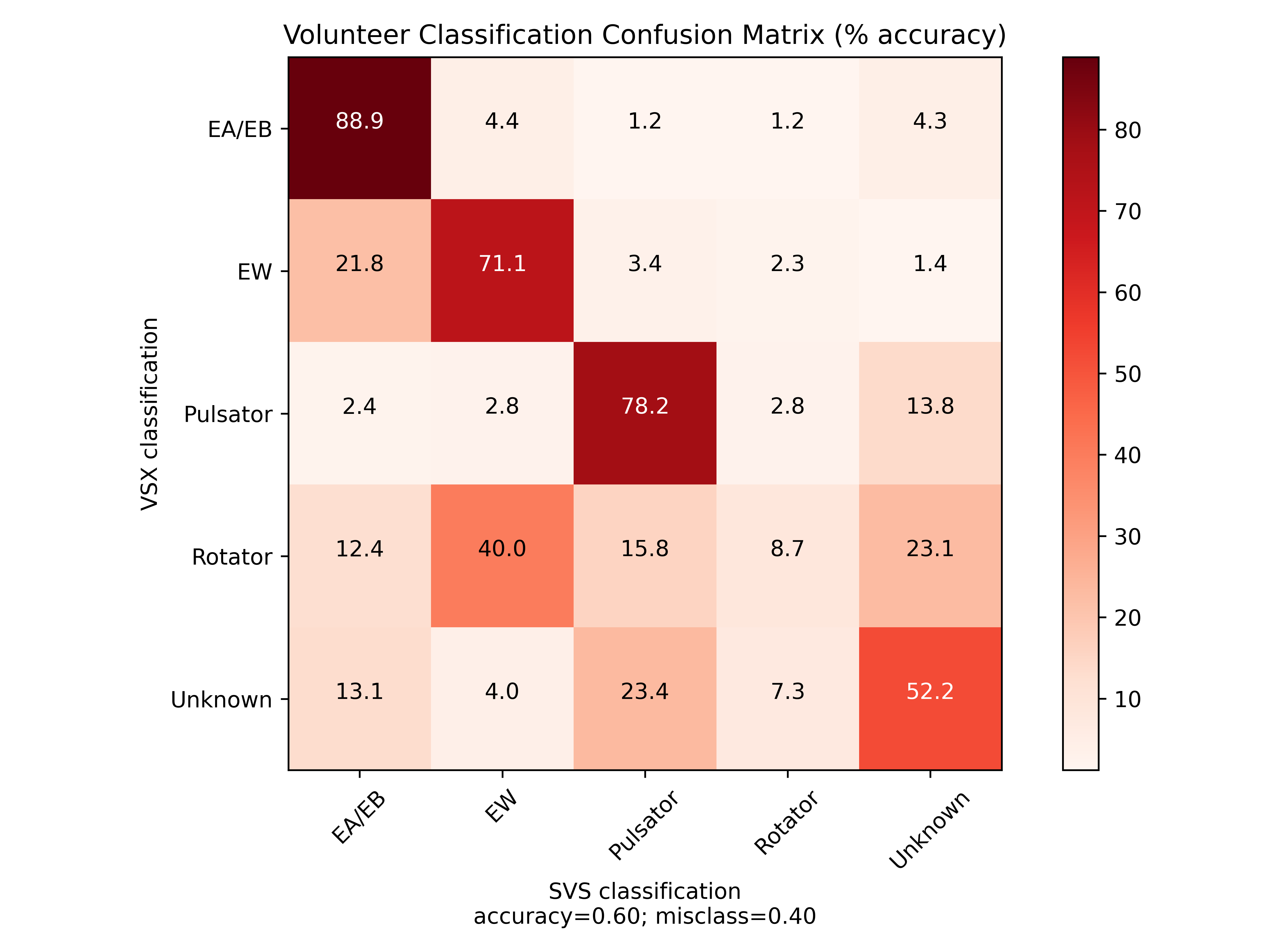}
 \caption{The confusion matrix for volunteer classifications compared to VSX classifications. The category of unknown for VSX contains semi-regular stars and stars classified as miscellaneous. We find an overall classification accuracy of 60 per cent.}
 \label{fig:confusion}
\end{figure}

To estimate the classification reliability, SVS classifications are compared existing variable classifications, such as VSX classifications or Gaia-DR2 variable types. Figure \ref{fig:confusion} shows the confusion matrix for volunteer classifications compared to the closest stellar variable within the VSX catalogue. While the SVS classification accuracy is high for binaries and pulsators, with $\sim$89 per cent of EA/EBs, $\sim$71 per cent of EWs, and $\sim$78 per cent of pulsators being correctly classified, rotators are a more challenging variable type with only $\sim$9 per cent of rotator classifications being "correct". The category of unknown easily categorised, but separating SVS classified objects into their corresponding classes from the VSX catalogue gives $\sim$24 per cent semi-regular variables, $\sim$23 per cent miscellaneous variables, and $\sim$15 per cent long period variables. Overall, we find a classification accuracy of 60 per cent for all variable types, excluding junk.

Too few SVS variables have a Gaia-DR2 variability component to undertake a similar full assessment, using the \textit{Gaia-DR2 variability results} catalogue containing 363,369 classifications of pulsators from Cepheids to Mira variables. Only 1 EA/EB and 8 EW type SVS variables are classified as pulsating stars in Gaia-DR2. Of the 273 pulsators (27 per cent of 1020 identified) in \textit{Gaia-DR2 variability results}, 9 are classified as Type I or II Cepheids, 9 are Mira variables, 17 are $\delta$ Scuti stars, and 238 are RR Lyrae stars. 81 rotators and 47 unknown variables are classified as pulsators in \textit{Gaia-DR2 variability results}.

This assessment gives a rudimentary estimate on the probability that different classes of variables are classified correctly. When combined with the \textit{SuperWASP periodicity catalogue} likelihood statistics, we can use this to give us a good idea of the correct period and variability type. It is most likely that incorrect classifications arise from two causes. Some variable types, especially EA/EB, can appear to be another variable type when folded at the wrong period. It is therefore important that we have a robust method of identifying the true period of an object which may have multiple detected periods, see Section \ref{sec:multiple}. The other dominant cause of incorrect classifications will mostly likely be human error, and non-specialists may miss some of the nuances of a light curve that indicate a certain variability type. But a cohort of non-specialist volunteers is by no means a bad thing, since the combination of people-power and multiple classifications means that an accurate consensus is usually reached. Feedback from citizen scientist volunteers also suggests that confusion can arise from the overlaid binned red line, especially in instances where the binned line appears to show a different variable type from the actual data, due to data drop-outs or spikes. At this stage of the project, it is not possible to remove or edit this binned line, but it is something to be aware of in the analysis of the resultant classifications, and use of labelled data in machine learning. Other issues may arise if volunteers skip the training available to them through the Zooniverse interface, forget the training, or find the training is not written in their first language. 

While highly unlikely, it is also possible that bots, spamming, or deliberate sabotage can influence the results. There are no in-built protections against this on the Zooniverse platform, so the only way of identifying "spam" classifications is by checking for a high number of classifications by the same user within an unrealistically short time-frame. All classifications were checked for a single user making multiple classifications per second and none were found. It is not possible to check this for users who are not logged in, so unexpected spikes in classifications ($>$100 classifications in $<$1 minute) were searched for. Only one spike in activity matching these parameters was detected by a single user, and their classifications were visually assessed by the authors and verified as non-spam.

Volunteer weightings have not yet been implemented in the classification pipeline, but will be an important part of the CNN, and will be used to improve classification reliability. We trialled two simple methods of calculating weightings: identifying overlap of classifications with "expert" or author classifications, and overlap with VSX classifications. With 6 possible variable types, a suitable number of classifications is needed for each variable type to calculate weightings. Unfortunately the overlap with "expert" classifications is too low to provide a conclusive weighting. Assessing against VSX, we take only those have made $>$100 classifications of each variable type, of which only 15 have an overlap of $>$100 with VSX, which also provides an inconclusive weighting system. Alternative methods will be explored in future work, for example through the use of individual volunteer confusion matrices, see Section \ref{sec:need_ML}.

\section{Results}
\label{sec:results}

\subsection{Overview}
\label{sec:overview}

\begin{figure}
 \includegraphics[width=\columnwidth]{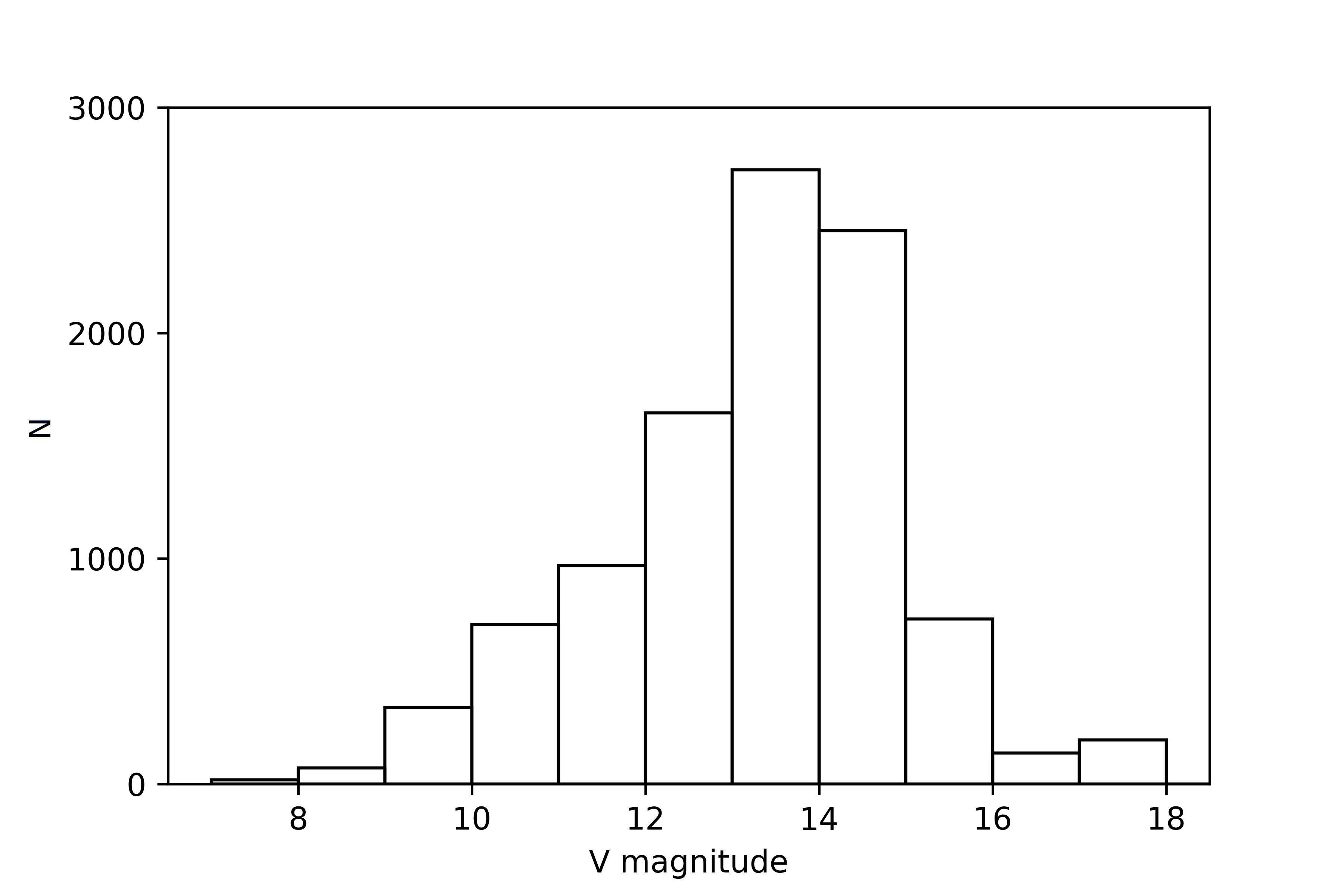}
 \caption{The distribution of NOMAD V magnitude of SVS stars with a variable type classification and correct period classification ranges between 8$\geq$V$\geq$18.}
 \label{fig:hist_mag}
\end{figure}

\begin{figure}
 \includegraphics[width=\columnwidth]{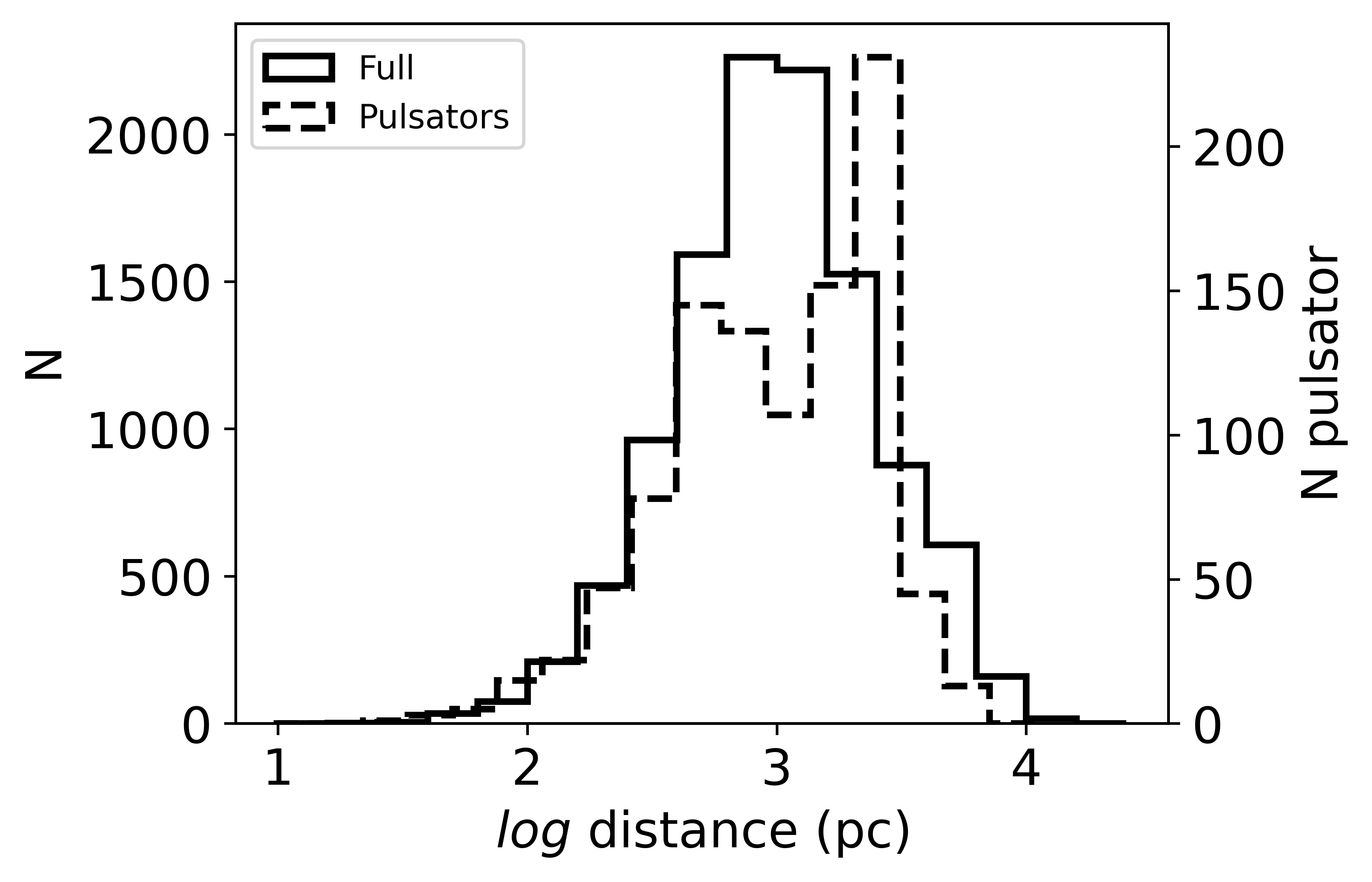}
 \caption{The distance (pc) distribution of SVS stars with a variable type classification and correct period classification. The full data set is shown in the solid line, while the pulsators are shown by the dashed line. Pulsators appear to have a different distribution to other variables.}
 \label{fig:distance}
\end{figure}

Volunteer classifications indicate that this first analysis consists primarily of junk classifications (66.6 per cent of all classifications), which are discarded. The remainder of the classifications are made up of EA/EB (5.3 per cent), EW (6.4 per cent), pulsators (4.5 per cent), rotators (9.9 per cent), and unknown (7.3 per cent). As previously identified, the classification accuracy of rotators is low so the true proportion will be lower than this figure indicates. Figure \ref{fig:hist_mag} shows the distribution of V band magnitudes ranging from approximately 8$\geq$V$\geq$15, with a number of fainter sources. Genuine faint sources can be detected by the longest SuperWASP exposures, but contamination by nearby stars can sometimes mimic faint sources, resulting in spurious detections. Figure \ref{fig:distance} shows the distribution of distances of these typically near-by stellar variables. Each variable type has a similar distribution, with the exception of pulsators, showing a peak in distance at $\sim$4800 pc, with a fainter average V magnitude of $\sim$13.8, likely due to a greater number of more distant stellar variables of this type. 

\begin{figure*}
 \includegraphics[width=2\columnwidth]{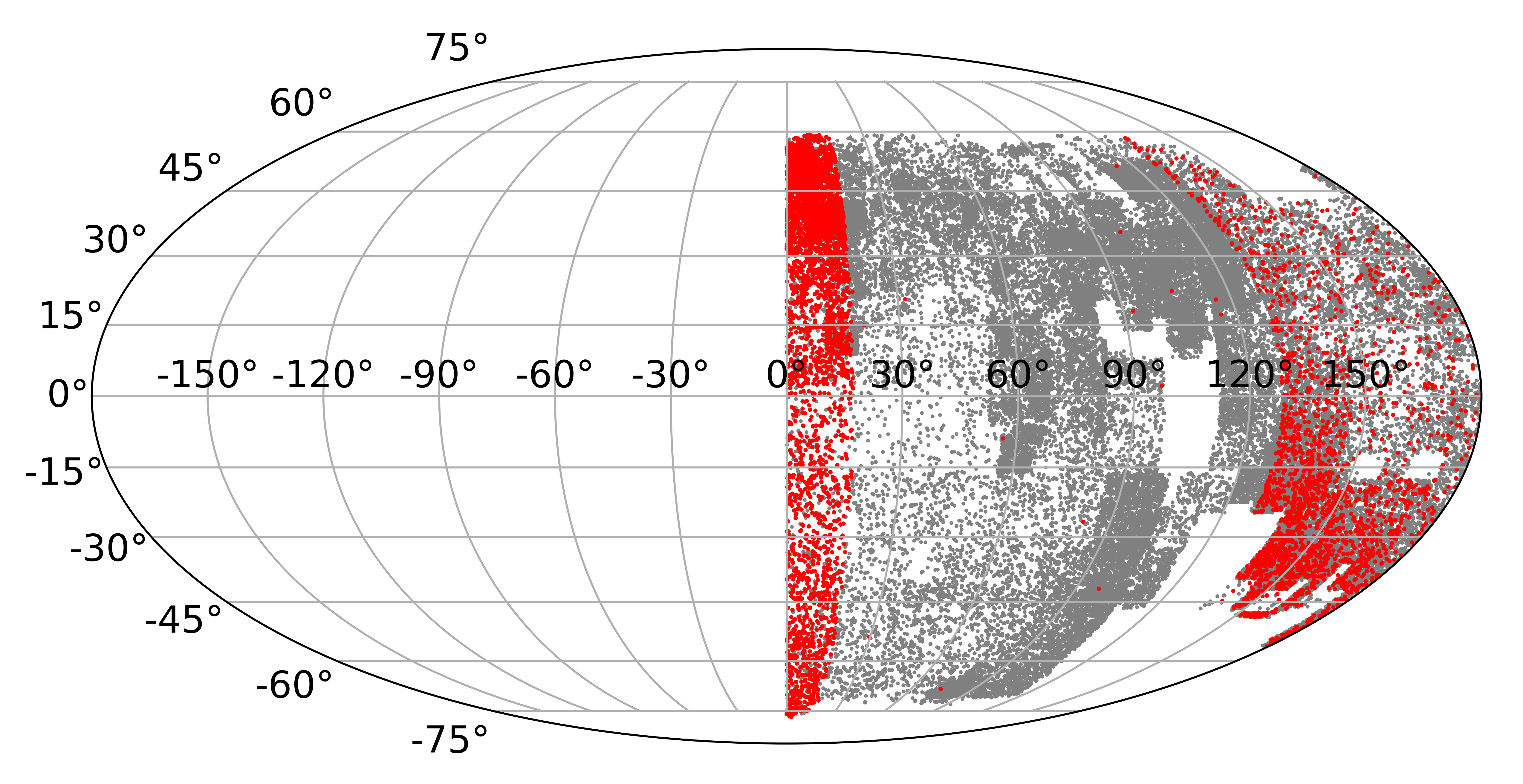}
 \caption{Map of SVS classifications. Red points indicate objects which have been retired from the live queue, grey points indicate objects which have received too few classifications for retirement. Classifications are not evenly distributed since only a few degrees of the sky are available to volunteers at any one time. As each data set is complete, more of the sky map will be filled.}
 \label{fig:sky_map}
\end{figure*}

The spatial distribution of the 568,739 unique object-period combinations is shown as a sky density plot in Figure \ref{fig:sky_map}. The classifications are not evenly distributed, since typically only a few degrees of sky are available for classification at any one time, and SuperWASP could not resolve objects in the dense regions of the Galactic Plane.

We have not yet accounted for the effects of interstellar extinction and reddening on magnitudes, colours, and variable classification. \citet{2018asasJ} make use of the reddening-free Wesenheit magnitudes (e.g. \citealt{1982madore,2018Lebzelter}), with Gaia DR2 and 2MASS passbands to improve variability classification for ASAS-SN, but do not account for the effects of extinction in colours. We aim to complete an analysis of the effect of both in future analyses of SVS classifications, making use of either the reddening-free Wesenheit magnitudes, the calculation of stellar extinction using the Binary and Stellar Evolution and Population Synthesis (BiSEPS) \citep{biseps} implementation of extinction given by \citet{drimmel}, or Gaia-DR2 reddening values and distances. Unlike ASAS-SN, magnitude and passband data does not feed into an automated classification pipeline, and our initial machine learning classification algorithm will not incorporate this data (Section \ref{sec:need_ML}). We expect that reddening would not be the cause of reclassification of the overarching variable types, however, for specific subsets of variable types (e.g. RR Lyrae stars), extinction correction may be necessary.

\subsection{New Variable Objects}
\label{sec:new}

\begin{table}
    \centering
    \begin{tabular}{c|c|c|c|c|c}
        \hline
        Type & EA/EB & EW & Pulsator & Rotator & Unknown \\
        \hline
        Number & 192 & 40 & 69 & 1,365 & 894 \\
        \hline
    \end{tabular}
    \caption{Previously unidentified stellar variables by variable type. There are significantly more variables classified as rotator or unknown. Stars classified as rotators are unlikely to be true rotators and may be binaries and pulsators folded at the wrong period, and unknown variables are likely to be junk, semi-regular or long period variables.}
    \label{tab:new}
\end{table}

\begin{table}
    \centering
    \begin{tabular}{c|c|c}
\hline
WASP ID                  & Type     & Period (days)  \\
\hline
1SWASPJ000005.14-755731.3 & EA/EB    & 4.30        \\
1SWASPJ000026.84+393855.6 & EA/EB    & 3.59       \\
1SWASPJ000028.05+041248.4 & EA/EB    & 4.69      \\
1SWASPJ000039.60-191306.0 & EA/EB    & 6.76       \\
1SWASPJ000047.05+353443.1 & EW       & 1.22      \\
1SWASPJ000054.70+544425.6 & EA/EB    & 3.19      \\
1SWASPJ000057.42-544520.1 & EA/EB    & 0.75      \\
1SWASPJ000059.84+094404.5 & EA/EB    & 0.65      \\
1SWASPJ000105.41-622920.6 & EA/EB    & 1.48       \\
1SWASPJ000132.23-051917.6 & Pulsator & 1.62       \\
1SWASPJ000132.66-091513.7 & EA/EB    & 4.19      \\
1SWASPJ000145.10+501843.4 & EA/EB    & 1.69      \\
1SWASPJ000149.26+061830.8 & EA/EB    & 0.32      \\
1SWASPJ000149.45-363918.1 & Pulsator & 0.64     \\
1SWASPJ000203.48-214746.0 & EA/EB    & 0.86     \\
1SWASPJ000315.40+495750.8 & EA/EB    & 3.65      \\
1SWASPJ000323.81+325049.7 & EA/EB    & 8.25      \\
1SWASPJ000343.16+465244.0 & Pulsator & 1.31      \\
1SWASPJ000353.60+043503.0 & EW       & 0.28      \\
1SWASPJ000410.77-525122.4 & EW       & 0.24      \\
\hline
\end{tabular}

    \caption{Sample from 301 previously unidentified stellar variables and related characteristics, not including rotators and unknown variables. The periods of each object have been assessed by the authors to correct for mis-classifications; whilst they have been corrected as much as possible, some periods remain best guesses. All periods have an uncertainty of $\pm$0.1 per cent. The full table, including rotators and unknown variables, can be found at \href{https://doi.org/10.5281/zenodo.4439383}{10.5281/zenodo.4439383}. 
    }
    \label{tab:new_full}
\end{table}

We expect SVS to classify many known stellar variables, and identify several previously unknown stellar variables. Previously known variables are identified by a 2 arcminute cross-match with the VSX catalogue (retrieved on 20 October 2020), which contains classifications of 2,105,377 variable stars from surveys including e.g. OGLE \citep{ogle}, ASAS \citep{ASAS}, ASAS-SN (\citealt{Shappee,Kochanek,2018asasJ}), ROTSE \citep{rotse}, NSVS \citep{nsvs}, ZTF \citep{Bellm}. A secondary cross-match is performed with catalogues from \citet{payne2013} containing 12,884 EAs, 5,226 EBs, and 2,875 EWs, and \citet{greer2017} containing 4,963 RR Lyrae stars.

To select potentially new variable stars, objects with a known classification and period are removed; objects that are flagged as variable, but which have no classification or period, are not removed. All new stellar variables were assessed by eye by the authors to verify the classification type and correctness of the period. Duplicated objects were removed and objects were reclassified as required. We caution that the subset of remaining rotator and unknown objects may still contain binaries and pulsators at the incorrect period, despite the best efforts of the authors to identify them. Through this process, we are left with 2,560 unique candidate new variables, shown in Table \ref{tab:new}.

\begin{figure}
 \includegraphics[width=\columnwidth]{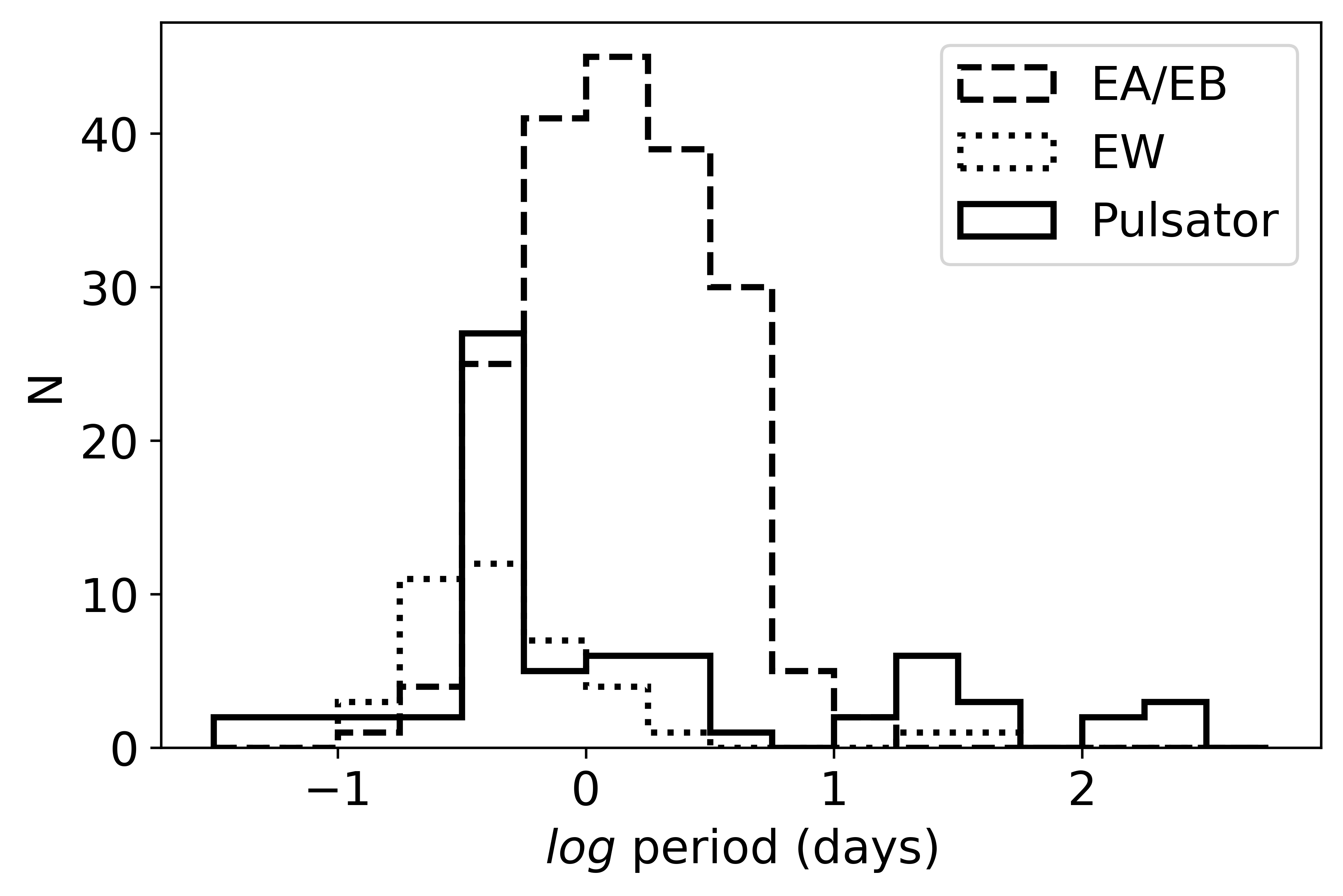}
 \caption{The distribution of period of newly identified stellar variables (EA/EB, EW, and pulsator) by variable type. EA/EBs are shown by the dashed line; EWs by the dotted line; pulsators by the solid line.}
 \label{fig:new_hist_period}
\end{figure}

Using this approach, we have identified 301 previously unknown variable stars, not including rotators and unknown variables, a selection of which are shown in Table \ref{tab:new_full}, with a period distribution shown in Figure \ref{fig:new_hist_period}. Of particular interest are a short period cutoff eclipsing binary (with two SuperWASP IDs: 1SWASPJ004003.56+501501.9 and 1SWASPJ004008.54+501455.6), new $\delta$ Scuti stars (Section \ref{sec:extreme}), and binaries displaying the O'Connell effect. Based on the low classification accuracy of rotators, we caution that new variables classified as rotators or unknown may not have the correct classification.

Excluding rotators and unknown variables, these new variables are typically bright (V$\sim$13) stars. It is likely that these objects have not been detected due to either surveys not yet having enough epochs to provide a variability classification (e.g. ASAS-SN), focus on the Galactic Plane or specific specific fields (e.g. Kepler, OGLE), or can only observe one hemisphere (ZTF). Assuming that 66 per cent of the 1.6 million light curves in SVS are junk, we estimate that on completion of SVS, $\sim$5,000 new EA/EB, EW, and pulsating stellar variables could be identified.

\subsection{Multiple Periods and Multiple Classifications}
\label{sec:multiple}

Stars displaying two or more real periodic modulations in their light curve are of great interest, and multiply periodic systems can act as stellar laboratories. Targets of interest are pulsating stars in eclipsing binary systems. There are detections of only $\sim$100 $\delta$ Scuti stars in eclipsing binaries \citep{2017Kahraman}, and there are very few RR Lyrae stars known in eclipsing binaries, and no known Galactic Cepheids in eclipsing binaries with orbital periods of less than 1 year \citep{2011evans}. 

\begin{figure*}
 \includegraphics[width=2\columnwidth]{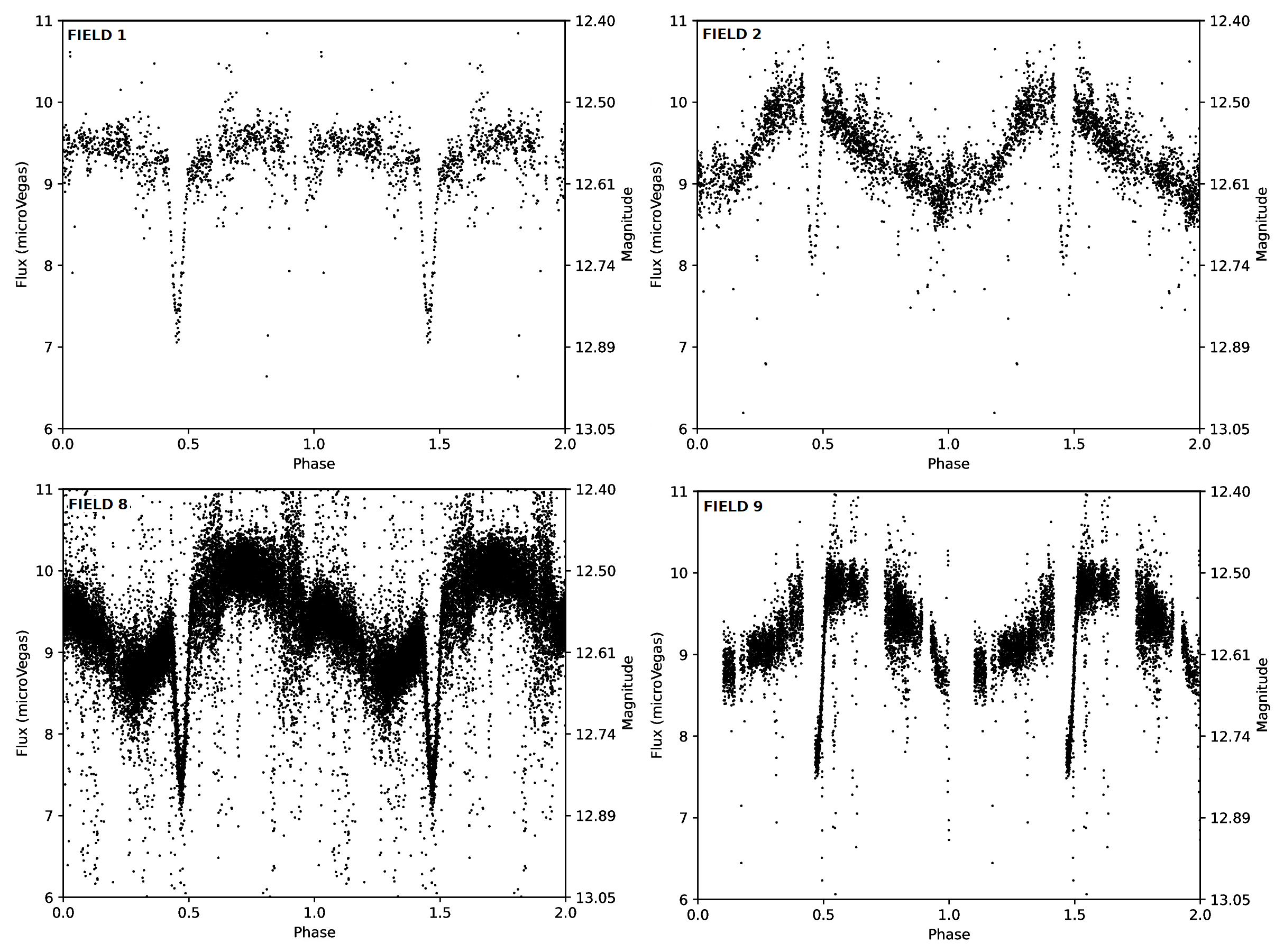}
 \caption{1SWASPJ004859.70+172328.1, an object with multiple EA/EB classifications, with a true period of 3.11 d. The midpoint of each frame is as follows: field 1 (August 2004), field 2 (August 2006), field 8 (October 2011), field 9 (December 2012).}
 \label{fig:eaeb_multi}
\end{figure*}

A search identified 1,202 multi-periodic systems, including 229 EA/EBs, 362 EWs, 100 pulsators, 441 rotators, and 70 unknowns. A visual inspection by the authors revealed that none are convincing multi-periodic systems, but instead are objects with aliases of the true period. Initially, 1SWASPJ004859.70+172328.1 appeared to have multiple correct EA/EB classifications. Further investigation found this object has a true period of 3.11 d, discounting the alias periods. However, this object has previously been identified as an eclipseless rotator (with a period of 3.11 d), but the SuperWASP light curves show a clear primary eclipse and shallow secondary eclipse, shown in Figure \ref{fig:eaeb_multi}. While the primary eclipse depth remains constant, the out of eclipse light curve changes significantly over the 8 years of observation, possibly due to a tidally locked star spot on one of the stellar components.

We are also interested in multi-classification systems. To identify such systems, we searched the SVS data set for subjects that have the same WASP ID but have multiple different, but by consensus "correct" period classifications. This search found 1,563 systems with 2 or more classifications, shown in Table \ref{tab:multi_class}. The classifications with the greatest overlap appear to be EA/EB and EW, and rotators with other classifications. Based on the low classification accuracy of rotators, we make the assumption that any multi-classification object in which one classification is rotator or unknown can be discounted as a true multiple classification. 

\begin{figure*}
 \includegraphics[width=2\columnwidth]{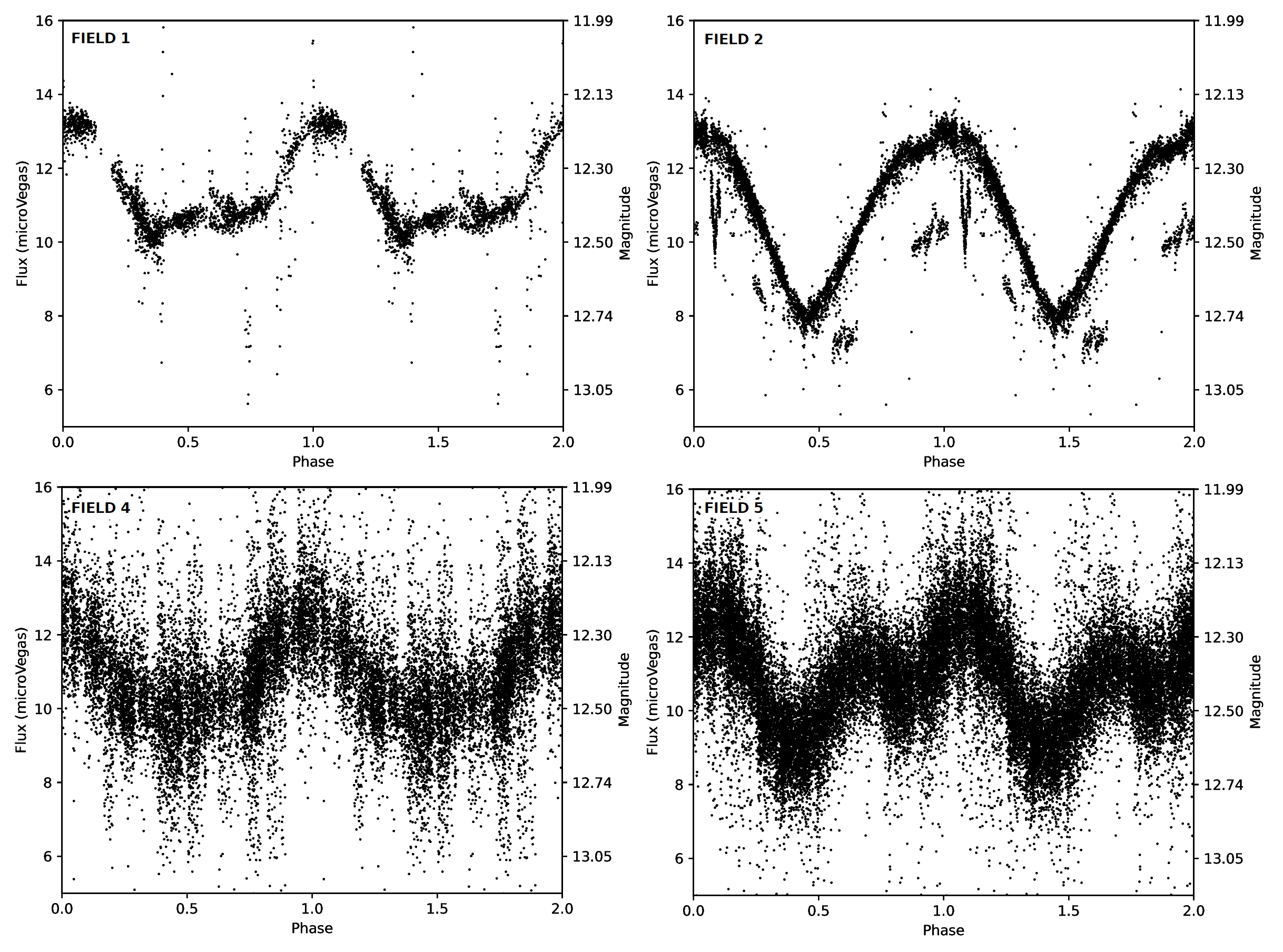}
 \caption{The light curve of 1SWASPJ000220.66-292933.8, classified by volunteers both as an EW with a period of 3.15 d and a pulsator with a period of 1.46 d. It has previously been classified as an eclipseless RS CVn and a non-periodic rotator. The midpoint of each frame is as follows: field 1 (September 2006), field 2 (September 2007), field 4 (August 2012), field 5 (September 2013).}
 \label{fig:rscvn}
\end{figure*}

Each of our candidate multi-classification systems were verified by eye (excluding rotators and unknown variables), ultimately yielding only apparently 1 real multi-classification system, 1SWASPJ000220.66-292933.8, shown in Figure \ref{fig:rscvn}. This object has both an EW and pulsator classification and SuperWASP periods of 3.15 d and 1.46 d respectively. On inspection, the EW classified light curve appears to be that of a RS Canum Venaticorum (RS CVn) binary. This object has a candidate RS CVn classification, with a period of 6.29 d or an eclipseless RS CVn classification with a period of 3.14 d in VSX. This object appears to have experienced significant surface spot coverage evolution over the 7 years of observations, and even hints at an eclipse in field 2.

\begin{figure}
 \includegraphics[width=\columnwidth]{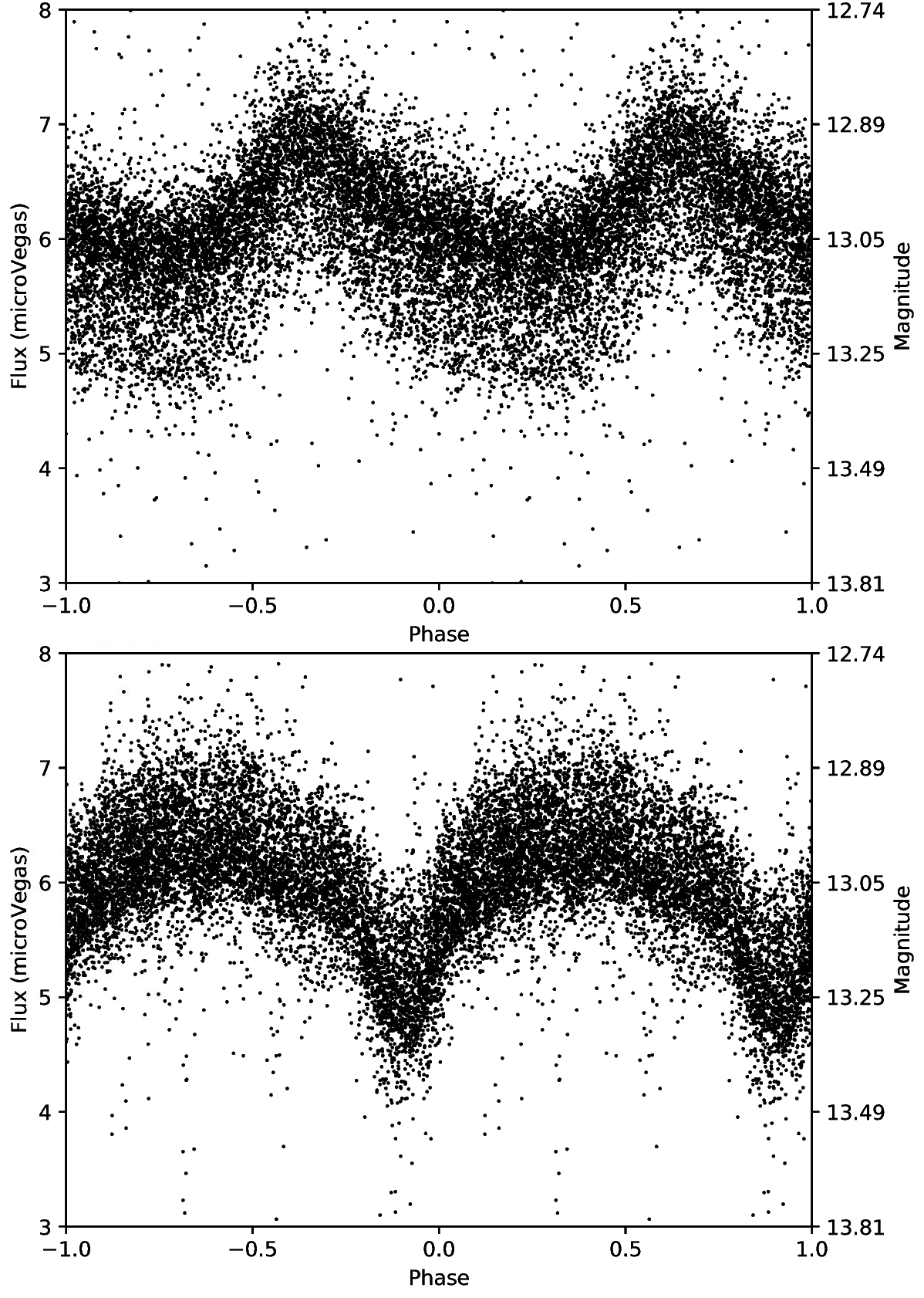}
 \caption{\textbf{Upper:} The $\delta$ Scuti star 1SWASPJ004811.15+473719.1 with a period of 1.9 hours. \textbf{Lower:} The EW-type eclipsing binary 1SWASPJ004810.36+473747.7 with a period of 18.7 hours (0.78 d). These objects were classified as the singular object 1SWASPJ004811.15+473719.1 with both an EW and a $\delta$ Scuti star in the same photometric aperture.}
 \label{fig:dscuti}
\end{figure}

Another object of particular interest was one which appeared to be a $\delta$ Scuti star in an eclipsing binary (1SWASPJ004811.15+473719.1), however this was found to be two separate systems, a binary (1SWASPJ004810.36+473747.7) and a $\delta$ Scuti star (1SWASPJ004811.15+473719.1), spatially separated by 30 arcseconds, shown in Fig \ref{fig:dscuti}.

\begin{table}
    \centering
    \begin{tabular}{c|c|c|c|c|c}
    \hline
         & EA/EB & EW & Rotator & Pulsator & Unknown \\
        \hline
        EA/EB & - & 246 & 128 & 5 & 75  \\
        EW & 246 & - & 716 & 16 & 46 \\
        Rotator & 128 & 716 & - & 99 & 202 \\
        Pulsator & 5 & 16 & 99 & - & 30 \\
        Unknown & 75 & 46 & 202 & 30 & - \\
        \hline
    \end{tabular}
    \caption{The number of light curves with multiple classifications per classification type. Rotators have the greatest overlap with other variable classifications, likely due to the low classification accuracy of rotators, and the high number of alias period light curves per rotator object.}
    \label{tab:multi_class}
\end{table}

\subsection{Extreme Variables}
\label{sec:extreme}

A valuable aspect of large catalogues of variable stars can be the identification of extremes of each class, i.e. those with extremely long or short periods, or extremely high or low amplitudes. SVS has the opportunity to increase the sample size of short period contact binaries, as well as identifying, for example, unusually long period contact binaries. For the full SVS data set, there are two peaks, at $\sim$0.3 days where we might expect to find short period binaries and aliases of binaries, and short period pulsators, and $\sim$30 days where we might expect to find semi-regular stars, currently classified as unknown. 

We explore extremes of each variable type using the following criteria as standard definitions of periods, and visually inspect light curves at the extremes of each period:

\begin{itemize}
    \item EA/EB: 0.3 d$\leq P \leq$10 d (e.g. \citealt{1995Stepien})
    \item EW: 0.22 d$\leq P \leq$1 d (e.g. \citealt{1992Rucinski})
    \item Pulsator: 0.3 d$\leq P \leq$8 d (e.g. \citealt{1912Leavitt,1979Breger,2006Matsunaga,2014Drake})
    \item Rotator: P$\geq$0.5 d (periods range from hours to months (e.g. \citealt{2013Nielsen})
    \item Unknown: N/A (semi-regular P $\geq$10 d) (e.g. \citealt{2009Soszy})
\end{itemize}

The class of pulsators has the widest range of possible periods, including $\delta$ Scuti ($\sim <$0.3d), RR Lyrae (0.44-0.82 d), Cepheid (with periods of weeks to months), Mira (P$\geq$100 d), and W Virginis (0.8 d$\leq P \leq$35 d). We chose a lower limit of P$\leq$0.3d to allow us to identify candidate $\delta$ Scuti and High Amplitude $\delta$ Scuti stars (HADS).

\begin{figure}
 \includegraphics[width=\columnwidth]{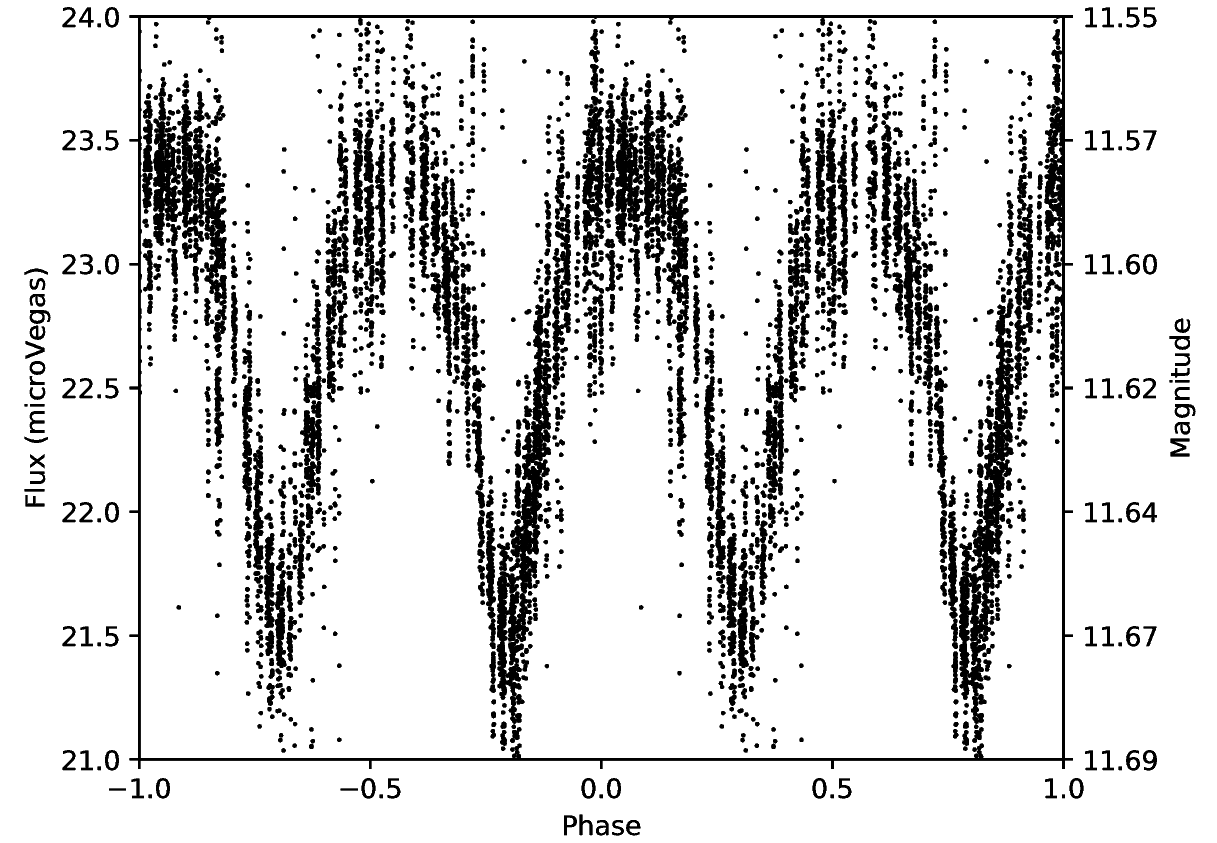}
 \caption{The first classification of a candidate near contact red giant eclipsing binary, 1SWASPJ000927.89+014542.1, with a period of 41.62 d, significantly longer than typical contact eclipsing binary periods.}
 \label{fig:ncrgeb}
\end{figure}

We have identified objects that appear to be long-period examples of near-contact eclipsing binary stars, with orbital periods of up to a month or more. To be in contact, or near contact, at such long periods requires the stellar components to be giants. Such objects have been proposed as the progenitors of red novae, but none have been conclusively identified pre-nova. The outbursts are believed to be due to stellar mergers, but only one progenitor of such an event has ever been studied, V1309 Sco, and that was only recognised retrospectively, after the merger occurred \citep{2011tylenda}. SVS volunteers have identified $\sim$10 candidates, with an example of one of these systems identified in SVS is shown in Figure \ref{fig:ncrgeb}. These candidate near-contact red giant eclipsing binaries are the subject of an ongoing follow-up campaign and the subject of an upcoming paper.

\begin{figure}
 \includegraphics[width=\columnwidth]{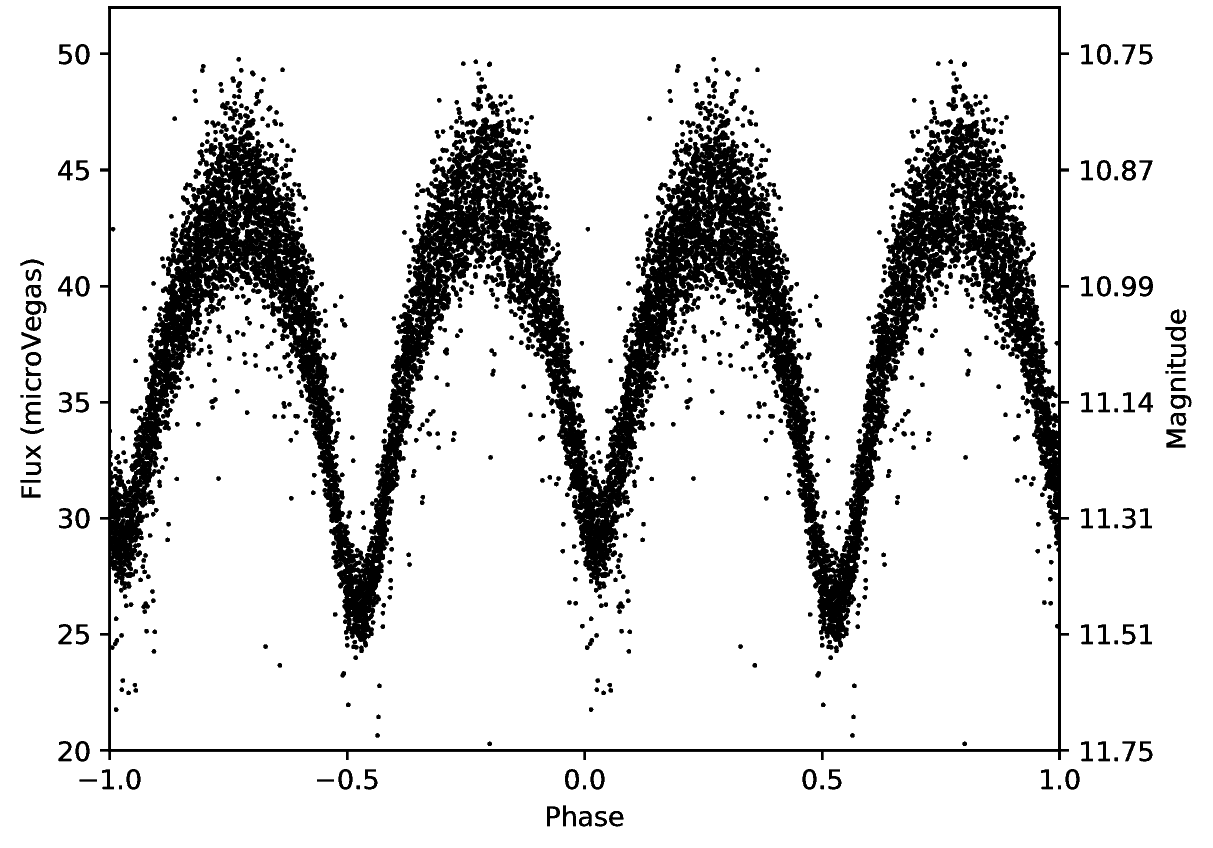}
 \caption{A newly identified EW type binary (both 1SWASPJ004003.56+501501.9 and 1SWASPJ004008.54+501455.6) with a period of 0.23 d, close to the short-period cutoff.}
 \label{fig:EW_short}
\end{figure}

We have also identified a new eclipsing binary (1SWASPJ004003.56+501501.9/1SWASPJ004008.54+501455.6) with a period of $\sim$0.23 days near the short-period cutoff of $\sim$0.22 days, shown in Figure \ref{fig:EW_short}. Such stars are of importance in the study of the evolution and structure of close binary systems.

\section{Discussion}
\label{sec:discussion}

In the full table of volunteer-classified light curves, we provide the SuperWASP ID, period (from the \textit{SuperWASP periodicity catalogue}), and best-guess variable type. We do not provide RA, Declination, or B, V, R magnitudes for any classified object. In most cases, there is only a single bright star in the photometric aperture and so this will usually be the source of the variability, so associations with other data are still possible most of the time. However. the large SuperWASP pixel size and possibility of contamination mean that we cannot confirm the association of a light curve with a specific stellar object without further follow up. We caution that anyone using this catalogue may need to confirm the variability type with their own follow up.

Although it is disappointing not to find many new multi-periodic or multi-classification systems at this stage, this analysis method can be applied to future analyses, especially for the identification of variables with evolving star spots. With a greater number of classifications, we expect to identify a significant number of extremely short and long period pulsators, including $\delta$ Scuti stars and Mira variables. Individual pulsator sub-types are not identified by citizen scientist volunteers, so would require the authors to visually inspect each pulsator light curve after making cuts using additional period, colour, and luminosity data. We also expect to identify more extreme binaries, including near-contact red giant eclipsing binaries, and binaries near the short-period cutoff. It is evident that if some form of machine learning is implemented, there may still be the need for some level of human interaction with multi-periodic and multi-classification systems to identify false positives.

We currently cannot estimate whether volunteer classifications have been biased. There is no identifying data on the image of each light curve, in an attempt to keep the classification task to a pattern matching exercise only. However, following the project launch, it was realised that some metadata for each light curve was visible to volunteers in the form of the SuperWASP ID. For volunteers who notice this, the ID gives information on the RA and Declination of each SuperWASP object, and hence the closest corresponding star from other catalogues. Subsequently, some users have used this ID to cross-match the light curve to existing classifications and surveys, using this knowledge to make a decision on the classification type. We do not have a way of identifying who has made use of this method and whether it can bias the results. Volunteer feedback has indicated that use of cross-matching has improved their knowledge of stellar variables and classification accuracy, and they value being able to investigate the light curves in more depth. 

To that end, as of November 2020, we have added links to external catalogues (CERiT, ASAS-SN, and Simbad) to the metadata which is visible only after a classification has been completed. It is not intended to be a tool to influence classifications, but it has been developed in order to allow interested volunteers to engage with the project further.

\subsection{The Future of SuperWASP Variable Stars}
\label{sec:future}

To successfully complete all classifications in SVS and make the results public, we are now working on implementing machine learning techniques and building a platform through which the results can be accessed.

\subsubsection{The Need for Machine Learning}
\label{sec:need_ML}

We estimate that at the current classification rate it will take at least 15 years to classify all 1.6 million light curves in SVS. To this extent, we are developing a novel method for classifying these phase-folded light curves to speed up the classification process, which is the subject of an upcoming paper. In this new method we will train a Convolutional Neural Network (CNN) on the same \textit{images} of phase-folded light curves as those presented to SVS volunteers. We will make use of the $>$1 million volunteer-generated classifications, or labels, to train the CNNs. We will run an initial CNN using volunteer-generated labels, then use expert classified light curves to calculate further volunteer confusion matrices, deriving \textit{fuzzy} labels and weighting classifications to improve reliability. We will then use a custom Zooniverse project to allow for expert bulk classification of CNN predictions, and retrain the CNN using expert classifications.

There is also the scope to use volunteer comments from the "Talk" forum section of SVS. It is possible for a volunteer to create a discussion page for each light curve, where they might "tag" or comment on it, giving a further classification type (i.e. while the SVS classification might be pulsator, a volunteer might comment "RR Lyrae" which indicates that the light curve is a pulsator sub-type). This forum potentially holds another significant source of labelled data which may be explored in future work.

\subsubsection{A New User Interface}
\label{sec:new_UI}

One of the key aims of SVS is to make the classified \textit{SuperWASP periodicity catalogue} light curves publicly available and to create the first catalogue of variable stars in the SuperWASP archive. We have begun work on a new user interface (UI), similar to WASP-DR1\footnote{\url{https://wasp.cerit-sc.cz/form}} and the ASAS-SN Catalogue of Variable Stars\footnote{\url{https://asas-sn.osu.edu/variables}}. 

This (UI) will take the form of a web portal, which will allow a user to easily and quickly search the classified light curves using a number of different parameters, including RA and Declination with a search radius, magnitude or flux, period, and variable type. A search of this UI will not only provide SuperWASP data and classifications, but also an automated cross-match to other catalogues, for example: SIMBAD, ASAS-SN, and VSX. Having selected an object, the user will be able to dynamically work with the data or download a FITS or CSV file. The dynamic interface will allow the user to fold the light curve at a different period, re-scale the plot, or convert between magnitude and flux, and more. This new UI will be updated with new SVS classifications or reclassifications every 6 months following its launch.

\section{Conclusions}
\label{sec:conclusions}

We present the preliminary results of the first analysis of the SuperWASP Variable Stars Zooniverse project, which consists of 1,025,750 classifications corresponding to 568,739 unique object-period combinations. Over 4,500 registered volunteers had engaged with the project between September 2018 and September 2020.

Each SuperWASP light curve has been classified by between 4 and 7 volunteers, classifying it as a broad type of stellar variable. We find that the majority (66.6 per cent) of classifications are junk and are therefore discarded, but the remainder (33.4 per cent) of the classifications corresponding to EA/EB, EW, pulsator, rotator, and unknown, are valuable for population studies and studies of unique stellar variables. We identified that variables with a rotational modulation are the most inconsistently classified by volunteers, with only $\sim$9 per cent of rotators being correctly classified, compared to $\sim$89 per cent of EA/EB type binaries. We caution that the classification of rotator should not be relied upon until there is a more reliable method of classification for this variable type. 

As a result of SVS, 301 new variable stars have been identified. Extrapolating to the wider data set, we would expect that $\sim$5,000 new variable stars could be identified on completion of this project. We have identified extreme period variables, including long period contact binaries, and eclipsing contact binaries near the short-period cutoff, and $\delta$ Scuti stars. This project has the potential to expand the catalogue of $\delta$ Scuti stars in eclipsing binaries, and discover the first Cepheids in eclipsing binaries (if they exist), as well as to identify multi-periodic Cepheids and RR Lyrae stars. The high number of false-positive multiply periodic and multi-classification light curves identified by volunteers indicates that an expert must complete the final stage of classification by eye for the most extreme and unusual light curves.

This analysis is not conclusive, but it demonstrates that SVS is successful in its aims of identifying unique and extreme variables, and identifying populations of stellar variables for further study. This analysis and methods will guide the project in future analyses of volunteer and machine learning classifications. We are now working on using citizen scientist classified data to train CNNs to speed up the classification process, however humans are still skilled at picking out the rare and unique objects, and generating labelled data. Both volunteer classified light curves and CNN classified light curves will feed into a new public user interface which is currently under development.

\textbf{Data Availability:} The full catalogue of 301 new variables discovered in SVS is available via Zenodo.

\section*{Acknowledgements}

We would like to recognise and thank the thousands of Zooniverse volunteers for their contribution to the SuperWASP Variable Stars project. We would also like to thank the Zooniverse team for their help in developing and maintaining the Zooniverse platform. The SuperWASP Variable Stars project was developed with the help of the ASTERICS Horizon2020 project. This publication uses data generated via the Zooniverse.org platform, development of which is funded by generous support, including a Global Impact Award from Google, and by a grant from the Alfred P. Sloan Foundation. This work was supported by the Science and Technology Facilities Council [grant number ST/P006760/1] through the DISCnet Centre for Doctoral Training. The SuperWASP project is currently funded and operated by Warwick University and Keele University, and was originally set up by Queen’s University Belfast, the Universities of Keele, St. Andrews and Leicester, the Open University, the ING, the IAC, SAAO and STFC. This research has made use of the International Variable Star Index (VSX) database, operated at AAVSO, Cambridge, Massachusetts, USA. This research has made use of the TOPCAT and STILTS software packages (written by Mark Taylor, University of Bristol). This research made use of the cross-match service provided by CDS, Strasbourg. This research has made use of the VizieR catalogue access tool, CDS, Strasbourg, France.



\bibliographystyle{mnras}
\bibliography{bib} 







\bsp	
\label{lastpage}
\end{document}